\documentstyle[aasms4,epsfig,psfig]{article}
\def\ltsima{$\; \buildrel < \over \simlt \;$}
\def\simlt{\lower.5ex\hbox{\ltsima}}

\def\simgt{\lower.5ex\hbox{>sima}}

\begin{document}

\title 
{Environmental mechanisms shaping the nature of dwarf spheroidal galaxies; the view of computer simulations}

\author 
{Lucio Mayer}

\smallskip
\begin{center}
Institute for Theoretical Physics,
University of Z\"urich, Winterthurerstrasse 190, 8057 Zurich, Switzerland
\end{center}

email:lmayer@physik.unizh.ch, lucio@phys.ethz.ch\\


\begin{abstract}
We review numerical work carried out over the 
last decade on the role of environmental mechanisms in shaping nature
of the faintest galaxies known, dwarf spheroidals (dSphs).
In particular we discuss a model in which dwarf spheroidal galaxies originated
from gas dominated, disky dwarfs that were accreted by massive galaxies
several billions of years ago. We assume the current cosmological
paradigm, the $\Lambda$CDM model, to define the structural properties of
the galaxies considered in the models as well as their orbital dynamics
and the effect of cosmic radiation fields.
We show how the combination of a tidally-induced 
morphological transformation, termed  tidal stirring, with mass loss
due to tidal  and ram pressure stripping aided by heating due to the cosmic ionizing
background can turn late-type dwarfs resembling present-day dIrrs
into ``classic' dSphs such as Draco and Fornax. High resolution 
numerical simulations incorporating all such mechanisms indeed produce
systems with the low angular momentum, mass-to-light ratios, stellar profiles,
luminosities, velocity dispersion profiles, and even the variety of
star formation histories exhibited by dSphs. The time of infall into
the primary halo is shown to be a key parameter. Dwarfs accreting at $z > 1$,
when the cosmic ultraviolet ionizing flux was much higher than today and
was thus able to keep the gas in the dwarfs warm and diffuse,
were rapidly stripped of their baryons via ram pressure and tidal forces,
producing very dark matter dominated objects with truncated star formation histories,
such as the Draco dSph. The low star formation efficiency expected in such low-metallicity 
objects prior to their infall was crucial for keeping their disks gas dominated until
stripping took over. {\it Therefore gas stripping along with inefficient star
formation provides a new feedback mechanism, alternative to 
photoevaporation or supernovae feedback, playing a crucial role in
dwarf galaxy  formation and evolution}. 
Dwarfs accreting at $z < 1$ were not completely deprived of their gas as heating
by the ultraviolet background has dropped in the meantime. They underwent bursts of star formation
at pericenter passages, producing systems with extended star formation histories
and lower mass-to-light ratios, such as Fornax. The transformation mechanism
naturally explains the existence of the morphology-density relation among 
dwarf galaxies, not only in the Local Group but also in generic group environments.
We discuss how hydrodynamical cosmological simulations are beginning to confirm the advocated 
transformation scenario.
Moreover, we examine the properties of the newly discovered ultra-faint dSphs in light
of this scenario and argue that they likely belong to a different population 
of lower mass dwarf satellites. These were mostly shaped by reionization rather
than by environmental mechanisms and are thus good candidates for being
``reionization fossils'' (see Ricotti, this Special Issue). We discuss
implications of the morphological transformation scenario
on the substructure problem, in particular on the transformation
between stellar velocity dispersion and halo circular velocity. Finally, we scrutinize
the various caveats in the current understanding of environmental effects as well as
other recent ideas on the origin of Local Group dSphs.

\end{abstract}

\section{Introduction}

Dwarf spheroidals (dSphs) in the Local Group are the faintest galaxies known. Including the
ultra-faint dwarf spheroidals discovered in the last few years, they span
luminosities in the range $-3 < M_B < -14$.
They are gas poor
and have pressure supported stellar components (Mateo 1998). Among them some
stopped forming stars about 10 Gyr ago and other have extended star formation
histories (Hernandez et al. 2000; Coleman \& de Jong 2008; Orban et al. 2008). 
They are typically clustered around the largest galaxy in a
group, although a few of them are found also at significantly larger distances
from the primary galaxy (Mateo 1998).
These properties of dSphs, while best studied and known in the Local Group
due to its proximity, are also typical of this class of galaxies in nearby groups and clusters.
Chiboucas, Karachentsev \& Tully (2009) have recently uncovered a population of dSphs in the
M81 group having a range of luminosities comparable to the LG dwarfs and which share
with the latter the same scaling relations between fundamental structural properties, such
as the relation between luminosity and effective radius. 
Recent studies of early-type dwarfs in clusters (Misgeld, Hilker
\& Mieske 2009) show that there
is overlap in the structural properties between the faint spheroidals in clusters and the classic dSphs in the LG, 
as shown by the size-luminosity and the surface brightness-luminosity
relations. Strong similarities between dSphs in the LG and in clusters are also found by Penny et al. (2009) in their study of the Perseus cluster.
Hence, LG dSphs can be considered as a representative laboratory for the study of dSphs in the Universe.
Mass loss from supernovae winds (Dekel \& Silk 1986), 
environmental mechanisms such as tidal and ram pressure stripping 
(Einasto et al. 1974; Faber \& Lin 1983) and suppression of gas 
accretion and/or photoevaporation during the reionization epoch 
(Bullock et al. 2000; Susa and Umemura 2004) have all been invoked to explain their origin and present-day properties.
In this paper we review the current status of theoretical models attempting to
explain dwarf spheroidals as the end product of environmentally induced morphological 
transformations from larger, gas-rich disky galaxies (Mayer et al. 2001a,b). The transformation
scenario is suggested by the fact that gas-rich, disky dwarf irregular galaxies 
(dIrrs) and dSphs
have similar stellar density profiles, close to exponential laws (Faber \& Lin 1983),
and obey the morphology-density relation whereby dIrrs are found in the outskirts of groups
while dSphs are mainly clustered around the main galaxy (Mateo 1998;Grebel 1999).
In hierarchical structure
formation subhalos nearer to the center of a virialized group halo or galaxy-sized halo were accreted earlier 
(Diemand et al. 2007); they were once beyond the edge of the primary halo
removed from its tidal influence.
We discuss how the combination
of tidal effects, ram pressure stripping and the cosmic ultraviolet background
radiation at high redshift may provide an explanation to both the similarities
and differences among the many known Local Group dwarf spheroidal galaxies.
We place our analysis in the context of the current paradigm for structure formation,
the $\Lambda$CDM model.
In this review we will always distinguish between  {\it classic} dSphs (Mateo 1998), those that were known
even before the advent of the SDSS deep photometric survey and have luminosities
$M_B < -7$, and the {\it ultra-faint} dwarfs discovered by the SDSS, whose luminosities
reach as low as $M_B  \sim -3$ (e.g. Koposov et al. 2008).
The classic dwarfs include Draco, Ursa
Minor, Sculptor, Sextans, Leo I and Leo II, Carina, Sagittarius, Cetus and Tucana
among the companions of the Milky Way, and the  many``And'' satellites of M31.
Based on the current measures of their internal stellar velocity dispersions
and on theoretical predictions on tidal mass loss of subhalos, we will argue that 
classic dwarf spheroidals were assembled in fairly massive
halos, with masses in the range $10^8 - 10^{10} M_{\odot}$. For these dwarfs reionization
played some role in depleting their baryonic content enhancing the effect of gas 
removal mechanisms
such as tidal stripping and tidal stirring as well as ram pressure stripping.
Such mechanisms were the major drivers of their evolution.
On the contrary,
the low halo masses of ultra-faint dwarfs make them likely candidates for
reionization fossils. The theoretical models that we present here are based on the results of
N-Body+hydrodynamical simulations of the detailed interaction of individual
dwarf galaxies with a primary host, as well as of hi-res cosmological simulations of 
Milky Way-sized galaxies.In the interaction simulations the model galaxies and the choice of the orbits are
consistent with the predictions of $\Lambda$CDM models 
(Mayer et al. 2002, Mayer 2005, Mayer et al. 2006, hereafter MA06).

\section{Current and past masses of dwarf spheroidals; why tides
are important }

Knowing the present and past mass of dSphs is crucial in order
to compute the effects of both environmental and internal mechanisms that
might affect their evolution. Moreover, determining the mass of dSphs, or,
equivalently, their circular velocity, plays a crucial role u
in the context of the missing satellites problem (Klypin et al. 1999;Moore et al. 1999).
Here we will concentrate on the effects that tides have on the evolution
of the mass and circular velocity of subhalos of Milky-Way sized
halos, and we will discuss how this folds in with the interpretation of 
the mass and circular velocity measured for observed dwarf spheroidals.

\begin{figure}
\hskip 2truecm
\epsfxsize=2truecm
\includegraphics[height=3.5in,width=3.5in,angle=0]{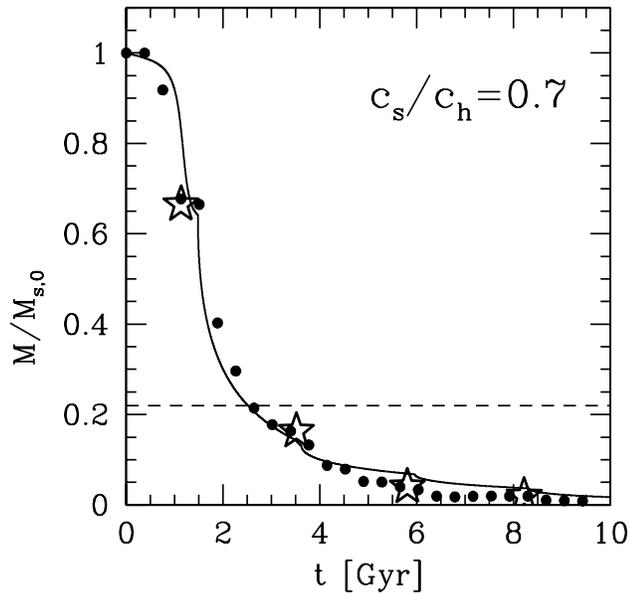}
  \caption{{\small Bound mass in units of the initial mass as a function of time, for a (dark matter-only) satellite moving on an 
eccentric orbit within a Milky Way halo
potential (see Taffoni et al. 2003 for details). The mass of the satellite is $1/50$ of the mass of the primary and both are modeled
with NFW halos.
The ratio between the concentration parameter of the halo and satellite is $cs/ch= 0.7$ , where $ch = 10$ is the concentration
of the primary halo. 
The symbols are the N-body data and the solid line shows the results of the semi-analytical model described in Taffoni et al. (2003),
that includes both tidal truncation and tidal shocks (see text). Stars identify each pericentre passage. 
The dashed line is the bound mass that would remain if only tidal truncation is applied.}}
\end{figure}

In cold dark matter models subhalos are accreted on highly eccentric orbits,
with a mean apocenter-to-pericenter ratio of 5-6 (Ghigna et al. 1998; Kravtsov
et al. 2004; Gill et al. 2004; Diemand et al.2007). Since the orbits are
eccentric, subhalos feel a time-dependent tidal force whose intensity is
strongest close to pericenter of their orbit.
The system undergoes an increase of internal kinetic energy as a result
of the tidal forcing, which modifies                                                  
the energy and mass profile of the subhalo, eventually
unbinding some of its stars or dark matter, namely producing tidal stripping.
The system will then attempt to reach a new virial equilibrium with a lower gravitational
binding energy, which in turn can make it more susceptible to the effect of
subsequent tidal shocks as its internal orbital/dynamical
time increases (see Binney \& Tremaine 1987). If a new equilibrium cannot be achieved the 
system will continue to expand, lower its gravitational binding energy and lose
mass, until it is eventually completely disrupted,
At low eccentricities the heating is reduced significantly,
until it becomes completely absent for a strictly curcular orbit (Gnedin et al. 1999).

Tidal heating in an extended mass distribution is most effective when
the internal dynamical time of the satellite is larger or at least commensurate with the
characteristic timescale associated with the galaxy's orbit (Aguilar \& White 1986;
Gnedin, Hernquist \& Ostriker 1999;
Taffoni et al. 2003). In this case it can be described using an extension of the 
impulsive approximation valid for a generic elliptical orbit of the satellite (Gnedin et al. 1999),
while the original impulsive approximation was derived for a point-like perturber and a target
(the satellite) moving on a straight-line path (Binney \& Tremaine 1987).
Following Gnedin et al. (1999) the
 expression for the heating term per unit mass for an extended perturber, but approximating the
orbit to a straight line path, is $<\Delta E> = (2 G M_0/V_{peri}{R_{peri}}^2) (r^2/3) \chi_{st}(R_{peri})$,
where $G$ is the gravitational constant, $M_0$ is the mass of the perturber, $R_{peri}$ and $V_{peri}$ are,
respectively, the pericenter distance and pericenter velocity, $r$ is the distance of a member star
from the center of the satellite, and $\chi_{st}$ is function
that accounts for the form of the mass distribution of the perturber (it would be
unity for a point mass). A more complex expression, but with similar scaling on the relevant
variables, can be obtained for elliptical orbits.
If, instead, the reverse condition applies to the two timescales
the system can readjust quickly to the perturbation and remain close to its original
equilibrium configuration. This second situation can be described analytically by
introducing the so-called adiabatic corrections. These are functions that reduce the
mean relative increase of internal energy $<\Delta E>/E$ derived via the impulsive approximation,
their value being proportional to the internal orbital frequency of the system. Based on the same
formalism it is easy to see that the innermost region of any system, that has the
highest density and thus the shortest dynamical time/orbital frequency, is also
the one that responds more adiabatically to the tidal perturbation.
The instantaneous orbital timescale, $t_{orb} =
2\pi R_{orb}/V_{orb}$, is shortest
at pericenter of the orbit, i.e. when $R_{orb} = R_{peri}$ and $V_{orb}= V_{peri}$,
which naturally implies the effect of the tidal perturbation will be stronger there.
Since the instantaneous orbital timescale, and thus the effect of the tidal perturbation,
varies the most when the galaxy approaches or leaves pericenter, this
is where the tidal force varies more strongly (recall the tidal force is relatvely short range, being 
proportional to $r^{-3}$, where $r$ is the distance from the center of the perturbing potential), hence the term "tidal shock".
If the orbit of the dwarf is circular there is no tidal shock.

Even before suffering the first tidal shock, a subhalo falling into the potential
of the primary halo for the first time would lose mass located 
at a radius $r$ larger than its instantaneous tidal radius $r_{t}$ (the tidal radius
of the subhalo decreases as it plunges closer to the center of the
primary halo). This other stripping mode is often termed "tidal truncation". There are thus two
mechanisms of tidal mass loss, tidal truncation, which typically shrinks
the subhalo to a radius of order $r_t$ evaluated at the pericenter
of its orbit, and tidal shocks (Taylor \& Babul 2001;Taffoni et al. 2003; Read et al. 2006a,b).
For subhalos with masses $< 1/50$ of the mass of the primary halo, as
true for most satellites of the Milky Way and M31 except perhaps
the Large Magellanic Cloud, dynamical friction has a negligible effect
(Colpi, Mayer \& Governato 1999),
thereby one can assume that their orbits did not change since infall. 
The growth in mass and size of the primary halo over time also has been
shown to have a negligible effect on the orbits of satellites in 
the case of a smoothly evolving analytical potential (Penarrubia et al. 2006),
although the clumpy and more erratic character of halo accretion and growth
observed in cosmological simulations may still have an impact on orbital dynamics
that should be investigated.  
With the assumption of a subhalo orbit constant over time
the tidal field of the primary near the pericenter of
orbit is fixed over time. Tidal truncation thus occurs only on the first orbit.
On the contrary, tidal shocks lead to continued mass loss on a fixed
orbit since each subsequent shock can heat the satellite and alter progressively
its internal structure. The process will cease to be effective only when the region
heated by the shock is dense enough to respond adiabatically to the external perturbation 
or when the satellite is completely disrupted y tides.
As shown in Figure 1,  analytical models using a combination of
tidal truncation on the first orbit and repeated tidal shocks describes
well the mass loss seen in numerical simulations (Taffoni et al. 2003).

\begin{figure}
\hskip 2truecm
\center
\epsfxsize=2truecm
\includegraphics[height=4in,width=3.5in,angle=270]{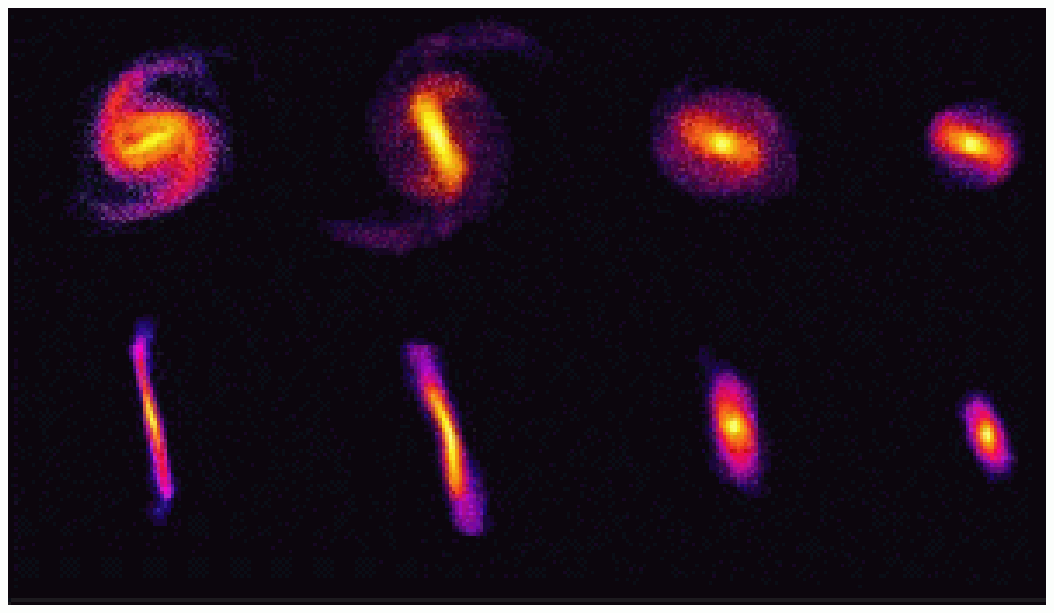}
\center
\includegraphics[height=2.3in,width=2.3in,angle=0]{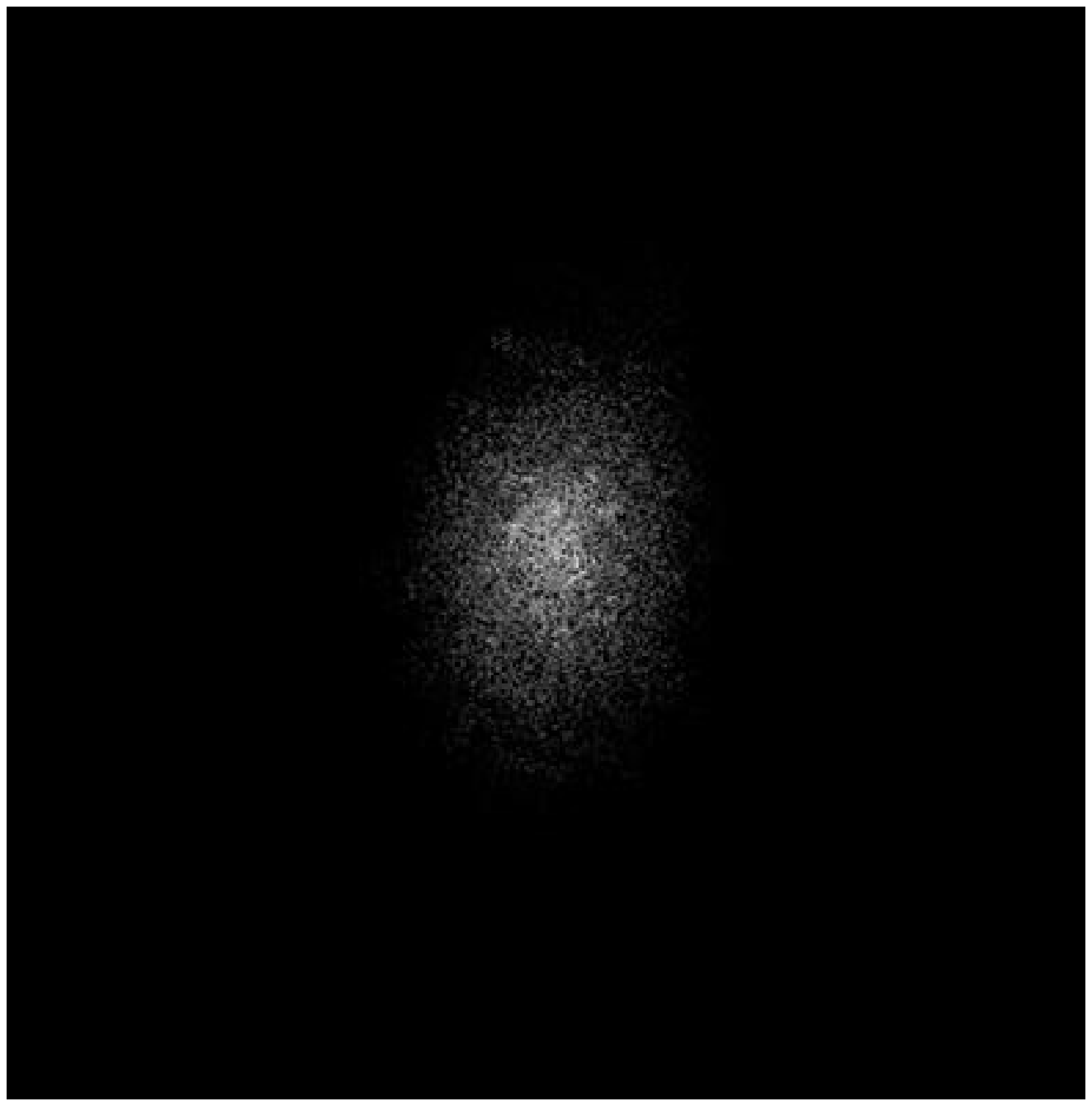}
\includegraphics[height=2.3in,width=2.3in,angle=0]{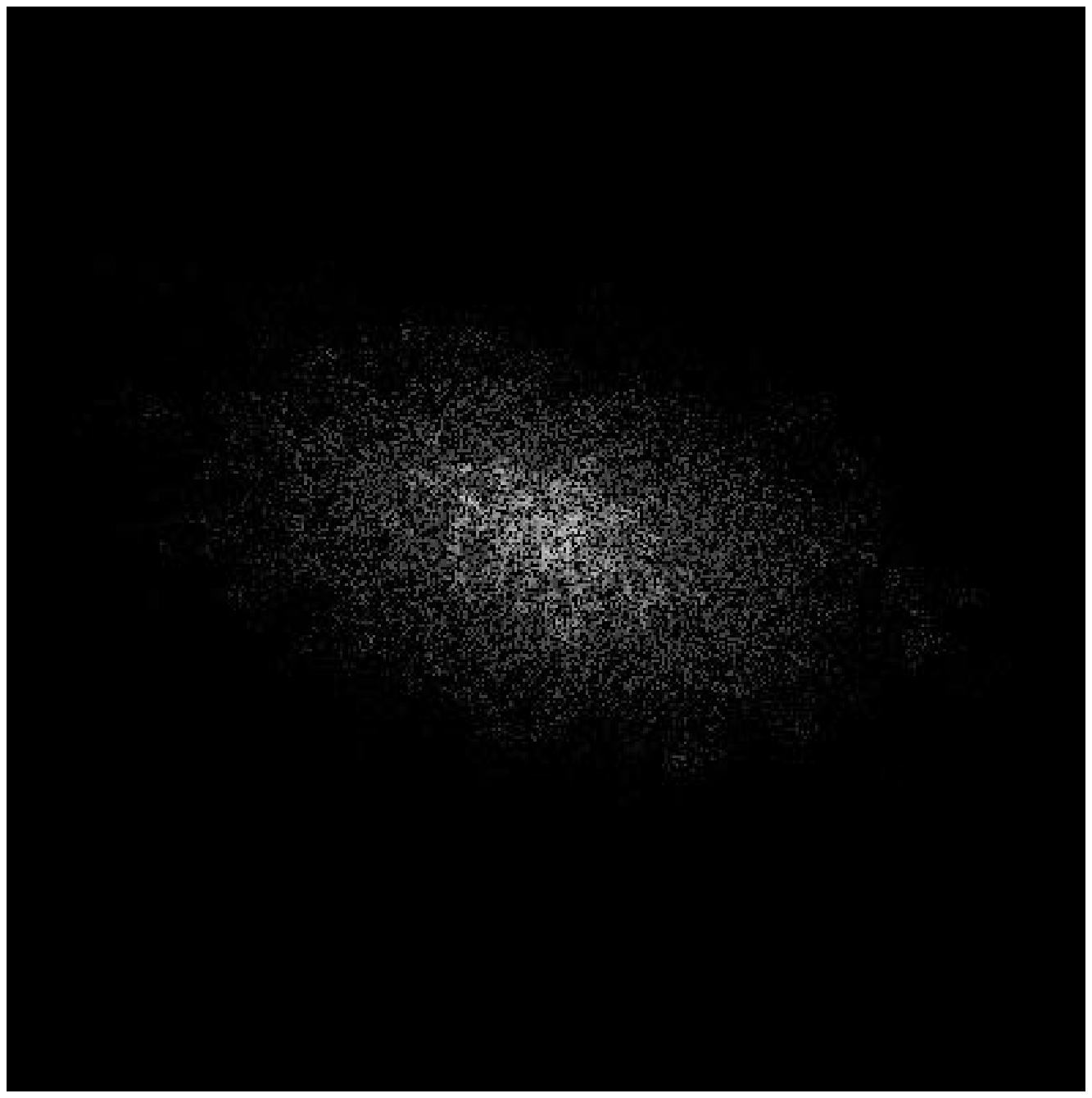}
  \caption{{\small {Top: Color-coded logarithmic density maps of a tidally stirred dwarf seen
face-on (i.e. with the line-of-sight perpendicular to the initial disk orientation, on
the top) and edge-on  (i.e. with the line-of-sight parallel to the initial disk
orientation, on the bottom). The time increases from the left to right, each snapshot
is separated by about half an orbital time (the orbital time is $\sim 2$ Gyr in this simulation), 
starting with the time corresponding to the first pericenter passage, when the
bar first appears. Boxes are 10 kpc on a side.
The edge-on view shows clearly the buckling of the bar occurring
on the second orbit, and the subsequent transformation into a diffuse spheroidal.
Bottom: gray-scale surface brightness maps of the remnant of the simulation shown
after 10 billion years of evolution (about 5 orbits). Two perpendicular viewing
projections are shown. The limiting B band  surface brightness shown in the images corresponds 
to $\mu_B = 32$ mag arcsec$^{-2}$ and a B band stellar mass-to-light ratio of 5 was adopted
for converting surface mass density into surface brightness (see Mayer t al. 2001b on 
models for the evolution of the luminosity and stellar mass-to-light ratio in the tidally stirred dwarf).
The faint simulated remnant resembles classic dSphs such as Draco or Carina in both apparent shape, 
luminosity and surface brightness distribution.}}}
\end{figure}

The amount of tidal mass loss of subhalos in numerical simulations can be
remarkable.  Kazantzidis et al. (2004a) have studied the disruption of
CDM subhalos within a Milky Way-sized primary with as many as $10^7$ particles. 
This approach has the advantage of 
allowing much higher resolution in subhalos relative to a fully cosmological
simulation that has to cope with modeling many objects simultaneously.
Tidal mass loss was found to be more
severe in less concentrated halos (where the concentration parameter is defined as
the ratio $c=r_{vir}/r_s$, where $r_{vir}$ is the virial radius of the subhalo
and $r_s$ is its scale radius, assuming an initial NFW profile)
because they respond more  impulsively
owing to their lower characteristic density and, correspondingly, longer
internal dynamical times.
Subhaloes typically lose
between $70 \%$ and $95 \%$ of their mass depending on orbits and concentration.
Correspondingly, their peak circular velocity, $V_{peak}$ decreases by a factor 2-3
over ten billion years of tidal mass loss. $V_{peak}$
reflects the inner mass distribution of the subhalo, at radii of order $r_s$, and is commonly used as a measure
of the characteristic subhalo mass throughout the literature. Its evolution over time 
is less strongly affected by tidal mass loss compared to the circular velocity of the subhalo 
at the virial radius.
Subhalos initialized with cuspy NFW profiles maintain
cuspy profiles with inner slopes close to $-1$ down to the force resolution, which in Kazantzidis et al. (2004a) was
smaller than the core radius of dSphs ($\sim 100$ pc) (see also Penarrubia et al. 2008a). This result
contradicts previous attempts to place dSphs within cored subhalos with the highest masses and $V_{peak}$
among the satellites' population in an attempt to solve the substructure problem (Stoehr et al. 2002 - see also 
Kravtsov, this Special Issue, and Penarrubia et al. 2008b).

Kravtsov et al. (2004)), using fully cosmological dark matter-only
simulations, found that $V_{peak}$ of the subhaloes drops on average by a
factor of $1.5-2$ in 10 Gyr of tidal evolution. 
More recently, Diemand et al. (2007) and
Madau et al. (2008a), using the high resolution Via Lactea simulation
of the formation of a Milky Way-sized halo, found that satellites
lose between 30 and 99\% of their pre-infall mass,
and that $V_{peak}$ decreases typically by a factor of $2-3$. This is consistent
with the recent results of Klimentowski et al. (2009) on the mass loss
of a ultra-high resolution subhalo with embedded stellar disk (see Figure 3, top
panel).
Quantitatively
similar results are obtained also
in individual subhalo-primary interaction simulations in which gasdynamics
with cooling and heating are included (Mayer 2005; Mayer et al. 2006, 2007, see also
section 5). Therefore there is now strong consensus on how tidal mass
loss affects subhalos, both qualitatively and quantitatively.

\begin{figure}
\hskip 2truecm
\epsfxsize=2truecm
\includegraphics[height=1.5in,width=6in]{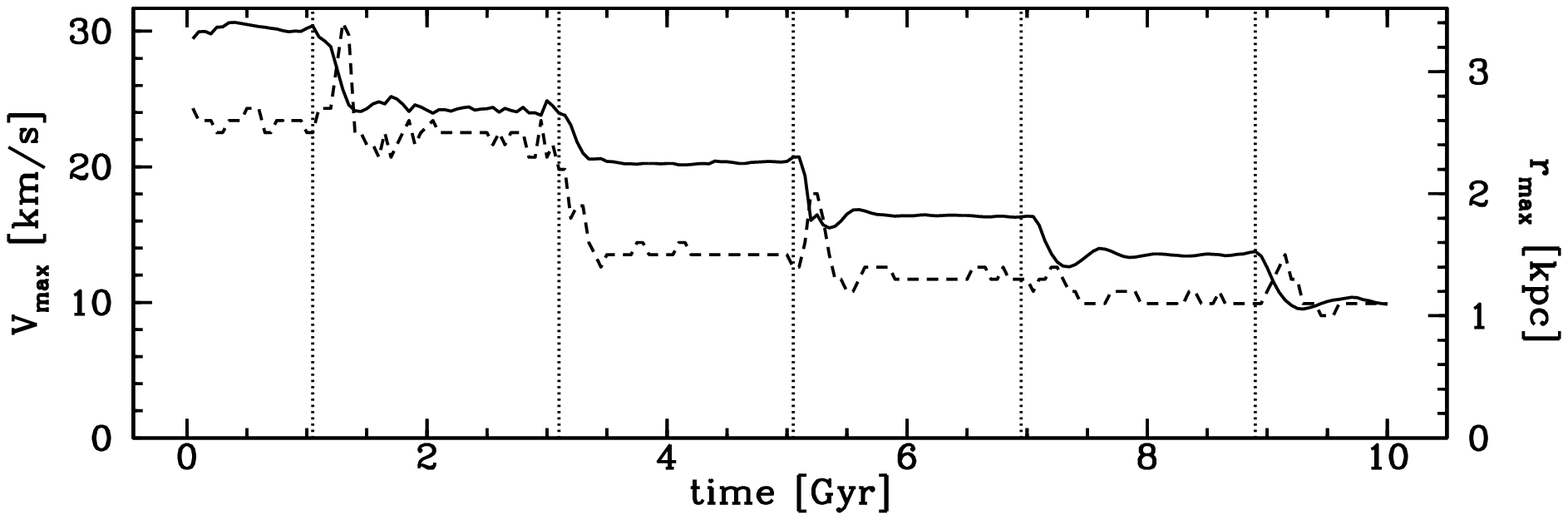}
\includegraphics[height=1.5in,width=6in]{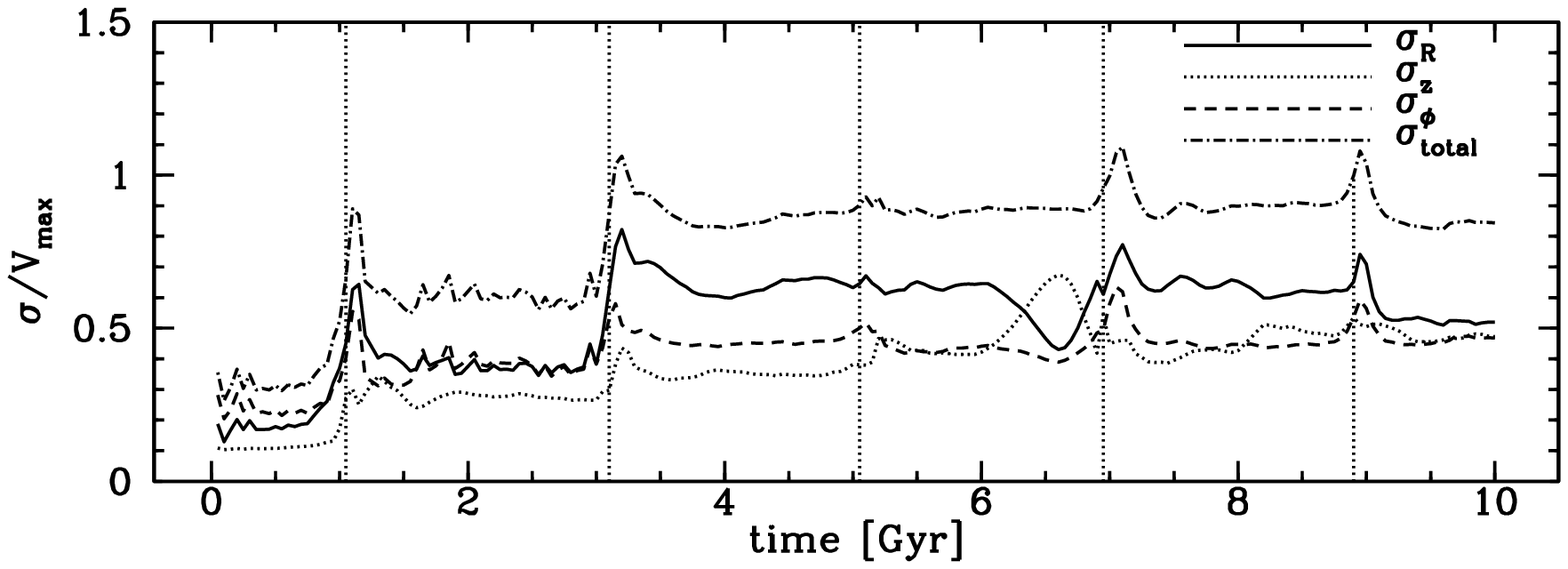}
\includegraphics[height=1.5in,width=6in]{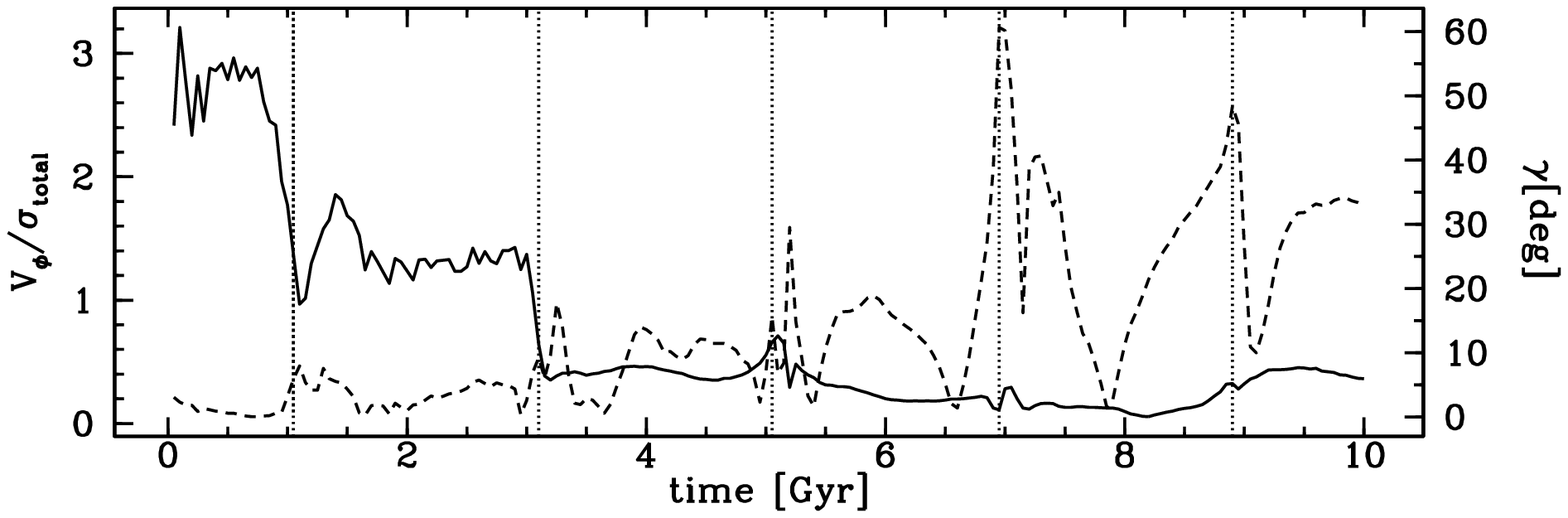}
\caption{
Evolution of structural parameters of a tidally stirred dwarf in a high resolution numerical simulation.
The plot show the results for one of the models
studied in Klimentowski et al. (2009a), whose initial disk was oriented at 90 degrees with respect to the orbital plane (this is a conservative case since other orientations leave almost no residual angular momentum - see text in section 3).
Upper panel: 
the maximum circular velocity (solid line, left axis) and the radius at which the maximum circular velocity 
occurs (dashed line, right axis). Middle panel: the velocity dispersion (total 3D, and 1D along three different cylindrcal axes) 
expressed in units of the maximum circular velocity (indicated as $V_{peak}$ in the text).
Lower panel: evolution of the ratio of the mean 
rotation velocity to the total (3D) velocity dispersion and the angle $\gamma$ between the total angular momentum vector of the 
stars and the orbital plane (dashed line, right axis).In all panels vertical dotted lines indicate pericentre passages
Loss of angular momentum due to tidally induced instabilities
and simultaneous increase of stellar velocity dispersion owing to tidal heating lead to a low $v/\sigma$ ratio comparable
to that of dSphs. See Klimentowski et al. (2009a) for details on how internal angular momentum is lost.} 
\end{figure}

Peak velocities for the best fit models to observed dSphs are at most a factor 
of 2 higher than those used in Moore et al. (1999) for subhalos, with the consequence that little  changes for the missing satellite problem.
Measurements of $V_{peak}$, both in simulated subhalos and real satellite galaxies of the
Local Group, are extremely important since they are used to express the missing satellite
problem, namely the excess in the number counts of subhalos at a given $V_{peak}$ compared
to the counts of observed dSph satellites of the Milky Way and M31 (Klypin et al. 1999; 
Moore et al. 1998, and see Kravtsov in this Special
Issue). However, in observed dSph satellites one measures stellar velocity dispersions
rather than the $V_{peak}$ of the subhalo. Going from stellar velocity dispersion to $V_{peak}$ requires assumptions
when the only information available from simulations is that provided by the dark matter component.
Typically (e.g Lokas 2002; Kazantzidis et al. 2004a) one uses the radial moment of the Jeans equations to compute
the relation between the circular velocity of the halo of the satellite and the velocity dispersion of the stars,
using the simulated subhalo dark matter profile but adopting
a simple functional form for a spherically symmetric stellar profile (typically a King model).
In this approach it is also necessary to assume a value for the
velocity anisotropy of the stellar system. As first shown by Lokas (2002), anisotropy
can play an important role in determining masses of dSphs from the data.
By varying assumptions in the Jeans modeling, including anisotropy, 
Kazantzidis et al. 2004 found that the observed stellar velocity dispersions of Draco
and Fornax ($\sigma \sim 10-14$ km/s) can be reproduced in subhaloes with $V_{peak} \sim 20-30$ km/s. 
Similar results were obtained by Zentner \& Bullock (2003).
Finally, this range is confirmed also by a more recent analysis that adopts a similar
methodology but takes advantage of the  cosmological Via Lactea simulation
(Strigari et al. 2007; Madau et al. 2008a).
The latter work finds values closer
to $15-20$ km/s for some of the other dSphs (e.g. Carina and Sextans). In Madau et al. (2008a)
a match between individual Local Group (LG) dwarf spheroidals and individual subhalos is achieved by 
considering information on the distances and spatial distribution of dwarfs.
In any case, the values of $V_{peak}$ obtained by these recent works for dSphs are
within a factor of 2 from those originally used in Klypin et al. (1999) and Moore et al. (1999)
to formulate the substructure problem
(note also that these two works did not use exactly the same conversion between circular
velocity and stellar velocity dispersion, see Kravtsov in this volume).

Recently, Klimentowski et al. 2009 have addressed the question of relating the observed line-of-sight 
stellar velocity dispersion $\sigma_{1D}$ of dSphs to the $V_{peak}$ of their halo
using simulations that, albeit not fully cosmological,
model both the baryonic and dark matter component of a disky 
dwarf subject to the tidal perturbation from a Milky Way-sized galaxy (see below).
They have found that the simple assumption that $Vc \sim \sqrt 3 \sigma_{1D}$, formally valid
for a tracer stellar population in an isotropic dark matter halo and used in Klypin et al. (1999), 
correctly describes the relation between the local stellar dispersion and the local halo circular velocity over 
a wide range of radii after 10 Gyr of evolution in the Milky Way potential (see Figure 3),
as explained in the next section (intermediate phases in which the stellar velocity dispersion tensor is 
more anisotropic produce temporary deviations from this relation).

In summary, the substantial simulation work carried out in the last few years suggests that
stellar velocity dispersions of present-day classic dSphs of the
MW and M31 in the range $\sigma \sim 7-13$ km/s ($7-20$ km/s if we include the bright dwarf elliptical satellites
of M31, NGC187 and NGC205) imply present-day halo $V_{peak}$
in the range $12-22$ km/s, and original halo $V_{peak}$ , i.e.
before infall and tidal mass loss,
in the range $24-44$ km/s (Kravtsov et al.
2004; Madau et al. 2008a). These last numbers are quite important. In fact,
with such relatively high values of the original $V_{peak}$
 photoevaporation by cosmic ultraviolet background at high
redshift and supernovae feedback cannot be what shaped the baryonic content
and nature of dwarf spheroidal as often advocated.
In fact photoevaporation would remove most of the gas only for $V_{peak} <
20$ km/s (Susa \& Umemura 2004). This is consistent
with the lack of a clear signature of the reionization epoch in the star
formation histories of dSphs (Grebel \& Gallagher 2004).
Second, substantial blow-out of the gas of dSphs due to supernovae winds cannot have occurred
since it requires $V_{peak} < 10$ km/s (e.g. MacLow \& Ferrara 1999; Read, Pontzen \& Viel 2006).
A caveat in the last line of reasoning is that in LCDM the progenitors of todays' dSphs formed by 
hierarchical accretion of smaller sublumps as any other galaxy and such
sublumps might have had a mass low enough to be affected by both reionization and
feedback. However, the old ages of the bulk of the stars in the majority of dwarf
spheroidals mean that their star formation epoch occurred more than 10 Gyr ago ($z > 2$). While 
dating of stars is non accurate enough to establish how much of the old stellar component
was formed before or after reionization, which occurred at $z > 6$, the cosmic UV background
was still close to its peak intensity at $z=2$ according to standard models 
(Haardt \& Madau 1996;2000), and yet  star formation was clearly happening.
This argues against an important effect of the cosmic UV background arising from reioinization,
perhaps owing to self-shielding (see Susa \& Umemura 2004). Note that,
conversely, there are examples of low mass LG dIrrs bearing a likely imprint of reionization in their
star formation history (e.g. Leo A, see Cole et al. 2007).

The same argument does not apply to the ultra-faint dwarf
spheroidals discovered in the last few years using deep photometry from the
Sloan Digital Sky Survey (SDSS) (e.g. Willman et al. 2005; Belokurov et al. 2008;
Koposov et al. 2008).
With a few exceptions, most of the extremely
faint dwarfs have very low velocity dispersions in the range $3-8$ km/s (Simon \& Geha 2007),
implying masses much lower than those of classic dSphs.
We will discuss them in section 8.

For the classic dSphs we advocated that physical processes associated with the environment, 
such as tidal forces and ram pressure stripping, were the major players in setting
their current properties and shaping their evolution.
The remainder of this review will be devoted to summarizing in quite some detail
the work done in this direction.
In the next two sections we will discuss how tidal interactions can completely
reshape dwarf galaxies, presenting a plausible model for the origin of dwarf
spheroidals in which such galaxies started out with a configuration and properties
similar to gas-rich dwarf irregulars and were subsequently transformed. We will
also discuss the issue of measuring the masses of dwarf spheroidals from the
observed stellar velocity dispersions, highlighting how tidal effects may produce
contaminated samples with a fraction of unbound stars projected along the line
of sight.

\begin{figure}
\hskip 2truecm
\epsfxsize=2truecm
\center
\includegraphics[height=4in,width=4in,angle=0]{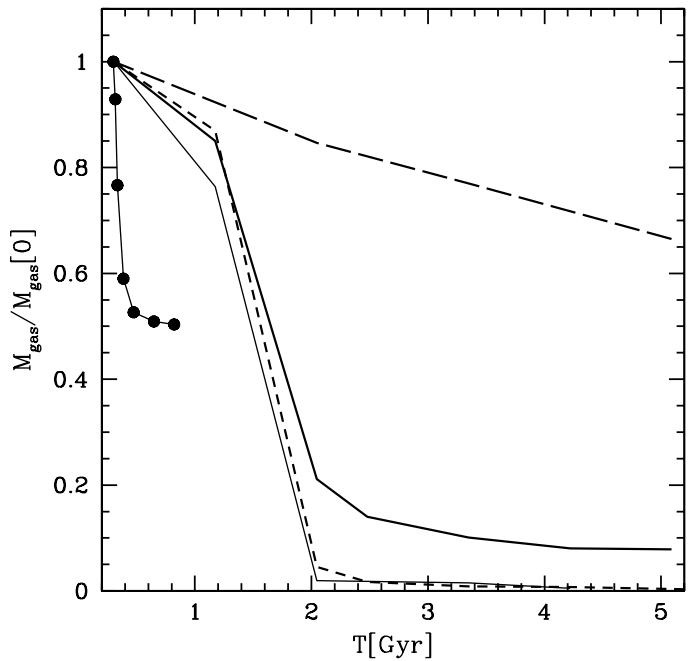}
  \caption{Time evolution of the bound gas mass for a disky dwarf
galaxy model(initial $V_{peak} =42$) km/s within a Milky-Way sized 
dark matter halo $(M_{vir} \sim 10^{12} M_{\odot})$ having a hot gaseous halo 
about 1\% of its total mass (see Mayer et al. 2006 for details on the modeling).
The dwarf's orbit has an apocenter of $150$ kpc and a pericenter of $30$ kpc.
The same model has been run with adiabatic gas conditions (thin solid
line), radiative cooling with no heating (thick solid line) and radiative
cooling plus heating by the cosmic ionizing UV background (dashed line), assuming that when the
simulation is started the time corresponds to $z=2$. The long-dashed line
shows the results for a run in which the hot gaseous halo of the Milky Way
was absent to isolate the effect of tides from that of ram pressure. 
Stripping from ram pressure only is instead shown by the thick solid with dots.
; in this run  the same dwarf model was evolved in a ``wind tunnel'', moving with a speed
comparable to the orbital velocity at pericenter in the other runs. In the ``wind tunnel''
the dwarf moves within a static gas distribution along a straight trajectory withe
specified velocity (periodic boundary conditions are used for the gaseous background),
but responds only to the ram pressure force. The
latter curve spans less than a Gyr of evolution since ram pressure stripping
saturates quickly without tidal mass loss.}
\end{figure}

\section{Morphological evolution of disky dwarfs into dSphs; tidal stirring}

Aside from the different gas content, three other facts
have to be considered when comparing dSphs and the other known type of dwarfs,
namely gas-rich, similarly faint dwarf  irregular galaxies (dIrrs) (see Mateo
1998 for a review on the subject)
First, except for the faintest among them all dIrrs in the Local
Group and nearby groups exhibit substantial rotation while dSphs
do not. Second, stellar profiles in the two classes of galaxies are similarly 
close to exponential. Third, a morphology density relation exists such that
dSphs are clustered around the primary galaxies while dIrrs are found at 
much larger distances from them.
Interestingly, Karachentsev (2005) 
finds that the fraction of dSphs over the total number of dwarfs in several 
nearby groups (including the Local Group)
decreases markedly at distances larger than about $250$ kpc from the
primary galaxies, comparable to the virial radius of haloes
hosting bright spiral galaxies in $\Lambda$CDM models. Clear trends
of dwarf galaxy morphology with environment for nearby groups (Centaurus A
and Sculptor) are also found by Bouchard, Da Costa \& Jerjen (2009).
The quantitative correlation found by Karachentsev (2005) strongly suggests,
although it does not prove, that dSphs are bound satellites of  bright spirals. It follows
then automatically that the environment provided by the extended halo of 
such massive hosts must be playing a crucial role in differentiating dIrrs from
dSphs. Of course there are outliers like the Local Group 
dSphs Cetus and Tucana,
located at more than 500 kpc from, respectively, M31 and the MW. However, 
cosmological simulations provide a potential explanation for those few distant dSphs; a few 
satellite halos on very plunging orbits can have apocenters exceeding the
virial radius of the primary (Ghigna et al. 1998), or can be on tighter orbits
and subsequently be ejected out to much larger distances via three-body interactions
(Sales et al. 2009). The outgoing radial velocity of Tucana further supports this possibility
(Fraternali et al. 2009). Overall, there are various effects that could
make the orbital dynamics of satellites much more complex than in our simulation
setup and accomodate such a case (see pag. 21-22).

\begin{figure}
\hskip 2truecm
\center
 \includegraphics[height=4in,width=4in,angle=0]{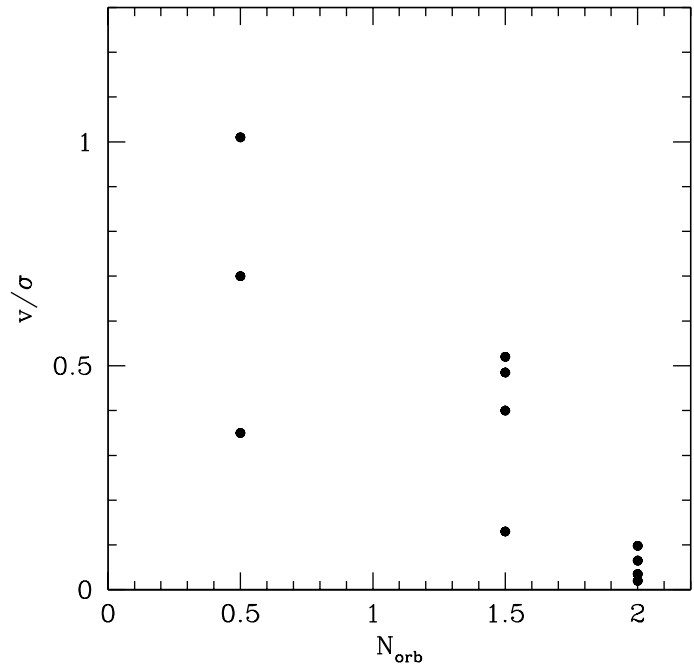}
 \caption{$v/\sigma$ as a function of the number of orbits for the satellites 
in the cosmological hydrodynamical simulation of Governato, Mayer et al. 
(2004).$v/\sigma$ is measured within the half mass radius of the dwarfs at the end
of the simulation, namely after 13 billion years of evolution.
The mean number of orbits for each of the three groups of dwarfs appears on the
horizontal axis.}
\end{figure}

Mayer et al. (2001a,b;2002), Mayer 2005, MA06, Mayer et al. 2007 and Klimentowski
et al. (2007,2009a) have shown that repeated tidal shocks at
pericenters of their orbits within the halo of a massive spiral can 
transform infalling disky dwarfs into objects resembling dSphs. The timescale
of the transformation is a few orbital times (several Gyr). The 
mechanism behind the transformation is tidally induced
non-axisymmetric instabilities of stellar disks  combined with impulsive tidal heating
of the stellar distribution. We recall that the time-dependent tidal field
varies on a timescale proportional to the orbital timescale of the galaxy.

The tidal forcing induces a perturbation to the internal energy and velocity 
distribution of the  system, which results into a perturbation of its mass and density
distribution as it attempts to acquire a new equilibrium configuration.
Through the density perturbaton tidal heating can thus excite 
modes in a self-gravitating system that would otherwise have a negligible amplitude.
If the induced velocity and density perturbation is considerable a strong mode might be
seeded and be subsequently amplified  by the self-gravity of the
system. An example of such induced mode amplification is the triggering of a bar instability,
which corresponds to an $m=2$ mode for the planar density field of the galaxy, of a warp, which
corresponds to an $m=1$ in the vertical density field (i.e. perpendicular to the
disk plane), or of a bending or buckling instability, which corresponds to an $m=2$ mode in the 
vertical density field (Raha et al. 1991). Numerical experiments, starting from the early 90s, 
have shown that the buckling instability can be triggered by a preceding bar instability.
A bar is supported by families of centrophilic stellar orbits, whose collective effect is
to increase the radial velocity dispersion along the plane of the galaxy. Raha et al. (1991)
first showed that above some critical value of the radial velocity anisotropy a system
would become unstable to the buckling instability - the whole disk bends above and below
and, as a result, stars acquire a larger
vertical velocity dispersion. The instability saturates
when the system has reached a new, nearly isotropic configuration in velocity space.
This picture has been confirmed by more recent simulations with much higher resolution
and physical complexity (Debatttista et al. 2006). As it reaches a new, nearly isotropic
configuration the system undergoes a transition from a cold disk into a spheroid. This
mechanism has indeed been proposed as a possible channel for the formation of bulges
starting from cold stellar disks (Kormendy 1993).

In simulations of disky dwarfs on eccentric orbits interacting with a Milky Way-sized halo the following
sequence of events is typically observed.
First, tidal shocks induce strong bar
instabilities in otherwise stable, light disks resembling those of 
present-day dIrrs. Second, the bar buckles due to the amplification
of vertical bending modes and turns into a spheroidal component in disks
with relatively high stellar surface density (Mayer et al. (2001a,b), or else subsequent shocks destroy the centrophilic
orbits supporting the bar which then loses its elongation and heats up into
a more  isotropic diffuse spheroid (Mayer et al. 2007; Klimentowski
et al. 2009a). An example of the first channel of transformation is shown in Figure 2 (top).
The second channel for the transformation is favoured in 
systems with lower mass, lower surface density disks; for these tidal heating
is particularly efficient because the dynamical response of the stellar system is
impulsive rather than adiabatic.
As shown in Mayer et al. 2007, the latter conditions likely apply to
the progenitors of the faintest classic 
dwarf spheroidals (dSphs) such as Draco, Sculptor or Leo I, or the And satellites of M31, 
while the bar-buckling sequence might have been more likely for the progenitors
of the brightest dSphs such as Fornax and Sagittarius, and even more so for those of the
bright dwarf elliptical (dEs) satellites of M31, such as NGC185 and NGC167 (but not
M32 whose unusually high surface density for a dwarf suggests it is the "threshed" core of a low
luminosity elliptical - see Bekki 2001). The bar-buckling sequence of instabilities also occurs for 
the progenitors of bright dwarf spheroidals and dwarf S0s in galaxy  clusters,
where both tidal stirring and harassment, i.e. repeated fly-bies with the brightest
cluster members, are at play (Mastropietro et al. 2005).
Tidal heating/tidal mass loss remove the disk outside the region that
goes bar unstable. Removal of the outer region, which contains most of the angular
momentum in the original disk, can be complete or partial, depending on the number
and strength of tidal shocks. Within this scenario, the LMC and SMC, and perhaps the two bright dEs of M31 which show 
fairly elliptical outer isophotes, are in earlier stages of the transformation, but
may also be sufficiently massive and dense to limit the tidal damage and never end up
into spheroidals (Mayer et al. 2001b).
Some of the faint dSphs, that show markedly elongated isophotes, such as Ursa Minor and some of the newly
discovered ultra-faint dwarfs (e.g. Bootes - see Belokurov et al.2006 - and Leo V - see Walker
et al. 2009), may owe such elongation
to a residual bar-like component rather than to tidal deformation.

\begin{figure}
\hskip 2truecm
\epsfxsize=2truecm
\center
\includegraphics[height=5.6in,width=5.6in,angle=0]{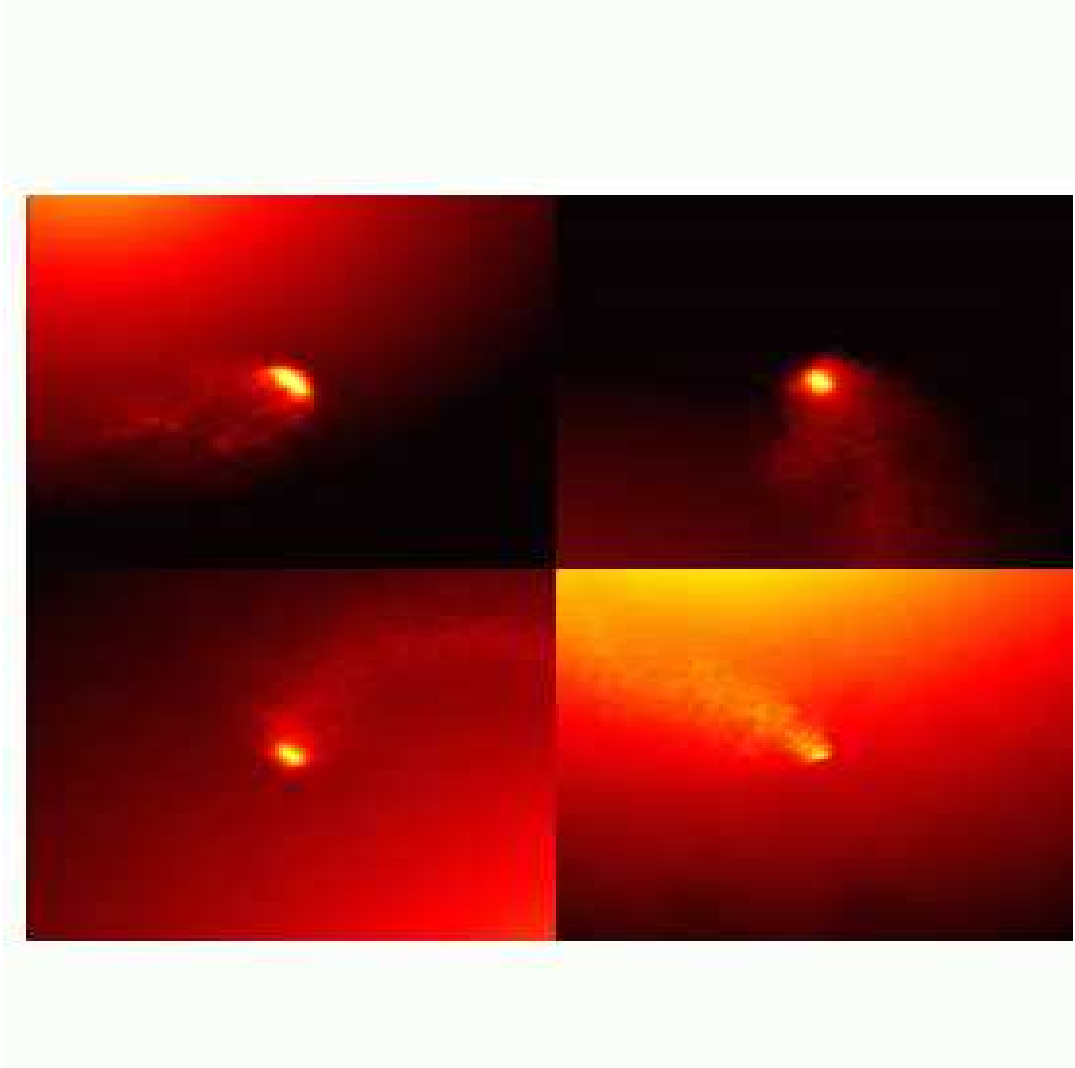}
\caption{Evolution of the gas-dominated disky dwarf model studied in
Mayer et al. (2007), which produces a dark matter dominated dwarf
resembling Draco after 10 Gyr of evolution. This N-Body + SPH simulation employed
millions of dark matter, gas and star particles. It included tidal mass loss due to a live
Milky Way halo, ram pressure stripping in a tenuous gaseous halo,
radiative cooling and a time-varying cosmic ionizing UV background consistent with
the Haardt \& Madau (2000) model. The snapshots show the color
coded logarithmic density maps (the brighter the color the higher the
density, with densities in the range $10^{-32} - 10^{-23}$ g cm$-3$)
for the first 2.5 Gyr of evolution. Boxes are 30 kpc on a side.
The dwarf begins falling
into the Milky Way halo on a typical eccentric cosmological orbit (apo/per$=5$ at the
beginning of the simulation, which corresponds to $z=2$).
At the first pericenter passage ($R_{peri} = 30$ kpc)  a prominent ram pressure tail is evident
(top left), and once the dwarf comes to first apocenter (top right) it
has lost already more than half of its gas. As it begins the second orbit
ram pressure stripping continues to remove gas (bottom left), 
until all gas is stripped at second pericenter (bottom right).}
\end{figure}

It is important to stress that tidal stirring operates independently
of tidal mass loss of the baryonic component. While systems that 
undergo substantial tidal mass loss are also effectively transformed,
the transformation can occur even when stellar tidal mass loss is marginal
(see for example the GR8 model presented in Mayer et al. 2001). This
implies that the detection of tidal tails in dSphs should not be used as
a direct way to test the feasibility of tidal stirring. Instead, the
deep structural analysis of dSphs and also of the so called transition dwarfs,
a class with intermediate properties between dSphs and dIrrs (Grebel 1999),
offers a much better test of the predictions of tidal stirring for the
different stages of the transformation mechanism. In clusters, the
discovery of disk-like features and bars in some dwarfs previously
classified as dSphs (Lisker et al. 2006) matches very well the expectations
of the tidal stirring and harassment models. While in the Milky Way halo
the main epoch of interaction between the primary and the satellite may
be over, as suggested by the the Sagittarius stream is the only
bright stellar stream present today in the Milky Way halo,
the study of other groups with more recent assembly histories may unfold
a population of satellites in the process of being transformed.

The remnants left behind by tidal stirring have nearly exponential profiles and very
low surface brightness (Figure 2). Tidal features typically lie at very low surface brightness
levels, below $32$ mag arcsec$^{-2}$ (Figure 2). 
The brightest among the remnants, 
those coming from progenitors
having relatively massive disks ($ > 10^8 M_{\odot}$) and high surface density disks, develop a central
steepening of the profile during the bar stage that persists even after buckling, and overall their
profile resembles that of the bright dwarf elliptical satellites of
M31, like NGC205 (Mayer et a. 2001a,b).
Most importantly, both "tidally heated" and "buckled" remnants have
a very low angular momentum, consistent with what is seen in 
dSphs and dEs, despite being descendants of disk-like systems.
The loss of angular momemtum is a by-product of the bar instability.
The bar sheds angular momentum outwards to the outer stellar disk
and dark matter, which are stripped after repeated shocks, thus
leaving behind a surviving stellar and dark matter core with very low rotation
, having a typical final  $v/\sigma < 0.5$.   The gradual loss of internal angular
momentum and decrease of $v/\sigma$ can be inspected in Figure 3.
Some residual rotation may be present at the end depending on the initial
structural properties and orbits of the dwarfs (Mayer et al. 2001). Klimentowski
et al. (2009a)  have also shown that for the same inital disky dwarf
model, the initial orientation of the disk angular momentum of the dwarf with respect
to the orbit also matters.
In orbits with high inclination, e.g. polar orbits,
$v/\sigma \sim 0.4$, perhaps consistent with what is found
in Leo I (Lokas et al. 2008), as opposed to indistiguishable from zero in coplanar or
moderately inclined orbits. How likely is one or the other configuration in CDM? The answer still
awaits us since it is likely that inclusion of the disk of the primary may
have played a significant role, for example by dragging satellites towards
its orbital plane and thus yielding a preferential bias towards a coplanar 
configuration (Read, Mayer et al. 2009). This would explain why most dSphs do not show sign of 
residual rotation.
In Lokas et al (2009) we analyzed the velocity structure of the remnants. In some cases we find radial
anisotropy in the velocity ellipsoid. This can be significant in remnants that have some residual rotation and are triaxial
due to a residua bar-like structure in the remnant, or can be negligible in those that have
been efficiently heated, turning into nearly isotropic spheroids with negligible rotation . 
In other words, when
tidal stirring produces a good match to a dSph, anisotropy in the kinematics is not important.

The transformation mechanism that we have studied requires the dwarf to be
on a bound orbit; for the mean apocenter/pericenter ratio found in
cosmological simulations, $\sim 5$, Mayer et al. (2001b) determine that
an apocenter distance comparable to the distance of Leo I (250 kpc), the
farthest dSph satellite of the MW, is a limiting case for producing a
dSph from a disky dwarf in less than 10 Gyr. However, this estimate
is very conservative and should be revised. In fact,in the meantime
we have learned that some satellites enter the primary system in sub-groups
and thereby might have 
already felt significant tidal perturbations before
entering the primaries (Kravtsov et al. 2004; D'Onghia \& Lake 2008).
Kravtsov et al. (2004) find that, at least in some cases, most of the
tidal heating in cosmological subhalos occurs before they enter
the primary halo. 
Furthermore, we have assumed that the only that could affect the orbit
of the dwarf within the primary halo is dynamical friction, and, moreover,
that its effect can be neglected for the typical low masses of subhalos.
While this would be a reasonable picture for dwarfs orbiting in a spherical, non-evolving primary halo, 
cold dark matter halos are triaxial and accrete mass via both smooth accretion
and mergers. Triaxial halos
are populated by families of chaotic orbits whose pericenters
can fluctuate over time. Yet, baryonic infall makes the potential
of the primary halo very close to spherical (Kazantzidis et al. 2004c),
whoch also agrees with current observational limits on the
shape of the MW halo (Koposov et al. 2009a), thus
suggesting that triaxiality should not not be an important effect. 
Instead, dwarfs infalling at very early epochs, when the primary halo
is still undergoing substantial growth via merging of relatively 
massive sub-groups, might undergo rapid changes in their orbits
as the global potential changes.
Another effect, which is seen to take place in simulations, is orbit scattering
via the interaction with another satellite, which can not only
modify the parameters of the orbit of the satellites but even cause
ejections when a three-body interaction takes place (Sales et al. 2007).

All simulations conducted to study the dwarf-primary interactions
neglect the effect of the disk of the primary. The disk of the Milky Way
was likely in place already at $z=1.5-2$, hence during most of the time
during which satellites were accreted and evolved within the primary halo.
While the tidal effect of the
disk is not expected to be important for typical pericenter distances of
satellites accreted below $z < 1$, because these should be larger than
20-30 kpc (see Mayer et al. 2001b), satellites accreting earlier had
smaller pericenters that would  graze the disk and could even be
quickly reduced by the additional dynamical friction provided by the
disk itself (Quinn \& Goodman 1986) if satellites accrete more frequently on
a prograde rather than retrograde orbit (Read, Mayer et al. 2009 
+ in prep.). D'Onghia et al. (2009a) find that disk shocking,
namely tidal shocks induced by passages through the disk, can
affect significantly the evolution of satellites having masses
$< 10^9 M_{\odot}$ and pericenters $< 20$ kpc, causing the
disruption of a fraction of them and thus playing a role in the
missing satellites problem (which nomally neglects disk shocking)
as well as in determining the radial distribution of satellites
inside primary halos.

This ``tidal stirring'' accounts for most of the similarities
and differences between dIrrs and dSphs, including the existence of the
morphology-density relation, by postulating progenitors of dSphs with
light, low surface brightness disks embedded in massive halos.
This, however, does not mean that such progenitors were identical
to present-day gas-rich dIrrs. Indeed, since they formed at high
$z$ they likely had assembly histories quite different from present-day
dIrrs, and thus their stellar populations and metallicities were 
different (this is an aspect that will be addressed by new, ongoing
simulations that start from a gas-rich dwarf galaxy formed self-consistently
in a cosmological simulation).
The next step would be to verify the model within a fully cosmological simulation,
and thus extend the analysis directly to the implications for the star formation
and metallicity histories of the dwarfs. Unfortunately, to date cosmological 
simulations with hydrodynamics have not allowed a robust analysis of the
structural evolution of satellites due to their limited resolution
(typically the force resolution is of order a few hundred parsecs - see
Mayer et al. 2008 for a review, so the tiniest dSphs like Draco are not resolved at all). 
Satellites, in addition, are too bright and dense in published cosmological
galaxy formation simulations due to numerical loss of angular momentum and
overcooling (Governato et al. 2007).
Searching for evidence of tidal stirring in these
simulations is thus quite hard, but it can be done in a qualitative sense for a 
few well resolved dwarfs. We selected the brightest satellites 
($M_B > - 14$) of the large spiral galaxy in the cosmological hydrodynamical 
simulation of Governato, Mayer et al. (2004) as well as similarly bright
dwarfs outside the virial radius of the same galaxy. Figure 5
shows their $v/\sigma$ within the effective radius versus 
the number of orbits performed within the main system. 
Clearly $v/\sigma$ correlates
well with the number of orbits as expected within the tidal stirring
scenario. We also found that the shape of the stellar components of the
dwarfs goes from more disky to more spheroidal with increasing number
of orbits. The absolute values of the $v/\sigma$ have to be taken with caution
because numerical two-body heating (e.g. Mayer 2004) is certainly an 
issue for these objects (halos have only a few thousand particles at 
this scale). However the trend is evident. A similar result, more recently,
has been obtained in a much higher resolution cosmological simulation of 
the formation of a massive early-type galaxy (Feldmann et al., in prep.) as
well as in newer, higher resolution simulations of the formation of a Milky-Way
sized galaxy (Callegari et al, in prep.).

\section{Tidal tails in disky dwarfs and line-of-sight stellar velocity dispersions}

As the dwarf is tidally perturbed and gradually transforms into a dwarf
spheroidal it also loses mass. In Mayer et al. (2002) tidal mass loss and
the formation of tidal tails in dark matter dominated, tidally stirred
disky dwarfs was studied using a range of initial models embedded
in an NFW halo and a range of cosmologically motivated orbits, i.e.
spanning and apocenter-to-pericenter ratio of 3 to 10.
The main conclusions were that (1) tidal tails projecting along the line
of sight could lead to an overestimate of the velocity dispersion (and
thus an overestimate of the mass-to-light ratio) by at most a factor
of 2 and (2) the most important contribution to the measured mass-to-light ratio is 
always the
intrinsic dark matter content of the dwarf. The corollary of the second
point is that the most dark matter dominated dwarfs have the faintest tails,
just lying at the detection limit of photometric surveys.
However, the role of projections effects, especially how likely
is that tidal tails lie perpendicular to the observer and thus can be
more easiy detected, was not clarified. Klimentowski et al. (2007) 
and Lokas et al. (2008) confirmed, in
a much more detailed study, that velocity dispersion profiles in dSphs
can be affected by contamination from unbound stars in the tails. While
the central velocity dispersion is not changed (the contamination from
unbound stars is statistically too weak since most of the stars projected
towards the inner core of the dwarfs are indeed bound), the outer velocity
dispersion can increase by nearly a factor of 2, with consequences on the
measurements of the total mass and mass-to-light ratio of the dwarf.
The same authors present a method of interloper removal to clean observed
samples from unbound stars contamination.
In Klimentowski et al. (2009b) tidal tails disky dwarfs were revisited.
The interesting conclusion is that
on typical cosmological orbits (apocenter/pericenter $\sim 5-6$)
tails point towards the observer for the largest fraction of an orbital revolution
of the dwarf within the Milky Way halo, especially where it spends most of the time,
namely near apocenter. This renders the detection of tails extremely difficult
and maximizes the enhancement of the line-of-sight velocity dispersion because
it implies that unbound stars belonging to the tails would be normally projected
along the line of sight, contaminating observed samples, although the effect is
complicated by variations in the intrinsic line-of-sight velocity dispersions
along the different axes of the dwarfs, see Klimentowski et al. 2007). The difficulty in
separating the tails from the bound core of the dwarfs due to this biased projection
effect may explain why in only a few of the dSphs, e.g. Carina 
(Munoz et al. 2006) and Ursa Minor (Martinez-Delgado et al. 2001), tails have been 
tentatively  detected. This effect adds to the most important requisite for tidal
tail detection, namely the need of an exceptional photometric technique (Munoz et
al. 2008).

\section{Gas stripping; ram pressure, tides and the cosmic ultraviolet radiation}

Mayer et al. (2001b) showed that tidal stripping alone cannot
produce the low gas fractions found in dSphs starting from a gas-rich
disky dwarf. Bar instabilities drive gas towards the center, making it tighyly bound.
Such inner gas distribution is not stripped by tides for the same reason why an inner stellar
core always survives in the tidally stirred dwarfs, namely because its binding energy is too high and the
response to tidal shocks is nearly adiabatic. 
Gas consumption by star formation does not solve the problem
if one adopts the mean star formation rates inferred
for dSphs (Hernandez et al. 2000; Orban et al. 2008); too much gas is left
even after billions of years (Mayer et al. 2001b).
In addition, gas turning into stars during the orbital evolution gives
rise to an extended star formation history, a feature that is seen
only in a subset of dSphs (Grebel 1999; Coleman \& de Jong 2008).
Therefore a mechanism that selectively removes the gas while leaving the stars
unaffected seems required.
Mayer et al. (2006) studied the combined effect of ram pressure and tidal
stripping. 
Ram pressure stripping occurs when the pressure force exerted by the ambient gaseous
medium exceeds the gravitational restoring force felt by the gas inside the dwarf.
The ram pressure is given by $P_{ram} = \rho_{gh} {V_{orb}}^2$, where $\rho_{gh}$ is the density of 
the ambient halo gas and  $V_{orb}$ is the orbital velocity of the galaxy
(Gunn \& Gott 1972). Assuming a constant ram pressure, there will be a characteristic
radius at which the ram pressure force $F_{rp} = P_{ram} S_{gd}$ (where $S_{gd}$ is
the surface area of the gas disk) is equal to the gravitational restoring force $F_{gr} (r) =
\nabla \phi (r)$, where $\phi (r)$ is the gravitational potential of the galaxy
The gravitational restoring force will indeed vary with radius since in galaxies
the potential is a function of radius. Since the gravitational restoring
force will increase with decreasing radius in a typical galactic potential
(e.g. see Abadi et a. 1999) the gas located at or outside the radius at which
$F_{rp} = F_{gr}$  will be stripped.

Mayer et al. (2006) constructed two-component models for the Milky Way halo in
which a dark matter halo consistent with the results of $\Lambda$CDM 
simulations
has an embedded diffuse gaseous component with a temperature of $\sim
10^6$ K and a density of about $\sim 10^{-4}$ atoms/cm$^3$ at 30 kpc from
the center, consistent with the values inferred from OVI absorption
measurements and the existence of the Magellanic Stream 
(Sembach et al. 2003). Dwarf galaxies are placed on eccentric orbits (apo/peri= 5-6)
with pericenters of 30 or 50 kpc. 
The initial models were gas-rich disky dwarfs ($M_{gas}/M_{stars} \ge 0.4$) 
It was found that ram pressure increases by a factor up to 10 the amount of
stripped gas mass compared to the case in which only tides are considered
(Figure 4). Ram pressure strongly depends on the depth of the
potential well of the dwarfs. While for dwarfs with $V_{peak} \le 30$ km/s
most of the gas content is easily removed, for more massive dwarfs 
the end result depends a lot on the orbit and on gas thermodynamics
(the effect of thermodynamics is seen in Figure 4, where we compare results
for an adiabatic run, a run with radiative cooling, and one with both radiative
cooling and the cosmic ionizing background).

For a galaxy moving on a non-circular orbit the ram pressure is not constant but
varies with time as both the galaxy's orbital velocity and the density of the
ambient medium change along its trajectory within the primary.
The pericenter distance sets the maximum strength of the
ram pressure force  since $P_{ram} = \rho_{gh} {V_{orb}}^2$, and both
$\rho_{gh}$ and $V_{orb}$ are maximum at pericenter. 
Moreover, there is another important time-dependent effect for galaxies on eccentric orbits.
Repeated tidal shocks lower progressively the depth of the potential well of the dwarf
at each pericenter passage, reducing the gravitational restoring force,
and thus allowing gas removal towards increasingly smaller
radii as time advances (this would be the case even if the ram pressure wind
was constant).
This explains why ram pressure stripping can continue over several
billion years (Figure 4) rather than saturating
after  the first pericenter passage.
However, when gas can cool down radiatively but
cannot be heated, the stripping process saturates after two orbits;
this is because the gas torqued by the bar flows towards the central region of the dwarf, increasing
its central density and rendering the response of the potential to tidal shocks more adiabatic
(see section 7). 

\begin{figure}
\hskip 2truecm
\center
\includegraphics[height=5in,width=3in,angle=0]{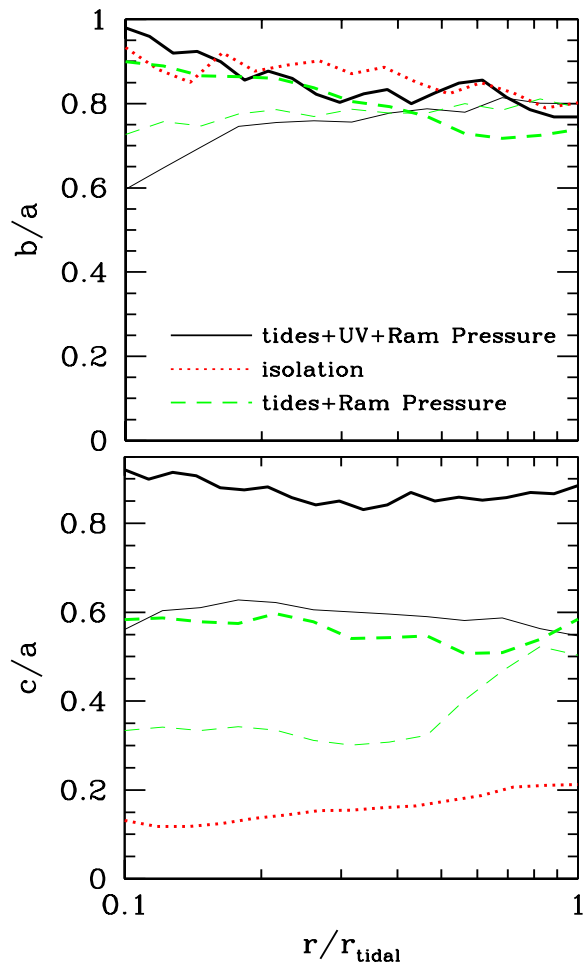}
\caption{Shape evolution of the stellar component. The upper panel shows 
the intermediate-to-major axis ratio, $b/a$, and the bottom panel presents 
the minor-to-major axis ratio, $c/a$. Solid lines correspond to the self-consistent
simulation in which tides, ram pressure stripping, and UV background were included
throughout the orbital evolution of the dwarf. Dotted lines correspond to the case
where the dwarf galaxy was extracted from the simulation just after the 
gas was removed and evolved in isolation.
Dashed lines show results of a test in which approximately 30\% of the
initial gas mass remains in the galaxy because the absence of
the UV background reduces significantly the effectiveness of ram pressure
stripping. Thick and thin lines correspond to 10 and 5 Gyr of orbital
evolution, respectively (for the extracted dwarf we only show the result after
10 Gyr). The shape of the stellar component changes gradually over time
and the presence of UV heating is essential for the shape transformation of thedwarf.}

\end{figure}

In principle stripping may still occur after the saturation induced by the bar infllow
via a mechanism that is not correctly captured by
the SPH simulations discussed 
throughout this paper, namely turbulent stripping by ablation due to Kelvin-Helmoltz instabilities
at the interface between the gas in the dwarf and the surrounding hot
diffuse halo (Murray et al. 1993;Agertz et al. 2007, but see how simple modifications of standard SPH can solve
the problem in Read, Hayfield \& Agertz 2009).
In hydrodynamical flows two shearing fluids moving with relative velocity $V_{rel}$ are subject to
an instability (the Kelvin-Helmoltz instability).
that tends to ablate their interface; a wave-like perturbation will grow proportionally
to the density contrast and relatve velocity of the fluids, but could be damped with increasing viscosity
or if an external force, such as a gravitational field, is applied to the system.

We can show that Kelvin-Helmoltz instabilities will not occur for dwarfs embedded
in a massive dark halo because of the stabilizing effect of the gravity of the halo
We neglect the effect of SPH artificial viscosity in the following calculation - this is
a conservative assumption since viscosity would stabilize the system further.  
Following existing analytical predictions (Murray et al. 1993)
we estimate that the critical value of the gravitational acceleration 
necessary to  stabilize the dwarf gas disk against Kelvin-Helmoltz in dwarf galaxy models with $V_{peak}= 40$ km/s 
used in Mayer et al. (2006) and Mayer et al. (2007) is $g_{cr} \sim 0.1$ initially, where $g_{cr} = {2 \pi V^2 / {D R_g}}$
($V \sim 200$ km/s is the average orbital velocity of the galaxy, $R_g = 0.5$ kpc is the scale length
of the gas component, and $D={\rho_g / {\rho_{gh}}}$ is the ratio between the density of the gaseous
disk at $R \sim R_g$ and the time averaged density of gaseous halo along the orbit of the dwarf, and for
simplicity we express the gravitational acceleration in the adimensional system of units of our simulations,
in which we assume $G=1$ for the gravitational constant $G$ ).
The self-gravity of the dwarf galaxy within $R_g$, which is mainly provided by the dark halo, is instead
$g  \sim 2.5 > g_{cr}$, where $g=GM_{halo}/{R_g}^2$. We note that $g > g_{cr}$ holds down to
about 1/5 of the disk scale length, and also for $R > R_g$. Hence the gaseous disk of the dwarf is
initially stable to Kelvin-Helmoltz instabilities.
However, on subsequent orbits the situation can change. For the simulation presented in Mayer et al. (2007)
after the second pericenter passage the density of the dwarf within $0.5$ kpc decreases by a factor of 4 as a result of the tidal
shocks, so that $g$ drops to $\sim 0.6$ at one
disk scale length, now being very close to $g_{cr}$ (the latter has also decreased in the meantime,
$g_{cr} \sim 0.3$, since the mean density of the gas has dropped).
The gas component of the dwarf is now only marginally stable to Kelvin-Helmoltz instabilities, but using the
formula to calculate the mass loss rate due to Kelvin-Helmoltz (Murray et al. 1993) for $R_g = 0.5$ kpc one derives
a characteristic stripping timescale still quite long, i.e. comparable to the Hubble time. 

There is then one more stripping mode to examine. Although formally 
standard SPH solves the equations for an inviscid fluid, {\it de facto} the necessity
of artificial viscosity renders the fluid model viscid. It thus makes sense to consider lamninar viscous stripping
as another possible stripping mechanism. The mass loss rate in this case is given by $\dot{M} \sim 12/2.8 \pi
{R_{g}}^2 V_{rel} \rho_{gh} (\lambda_{max} / R_g){(c_1/V_{rel})}$ (Murray et al. 1993, Nulsen 1982), where $\lambda_{max}$
is the longest wavelength not stabilized by gravity.  Assuming that, at second pericenter,
the longest unstable wavelength is $\lambda_{max} \sim 0.5$ kpc, one obtains a laminar viscous timescale
of $\sim 2$ Gyr for the typical ambient gas densities and relative velocities in our simulations ($V_{rel} = V_{orb}$).  
The gas is indeed completely stripped $\sim 1$  Gyr later in cases in which the gas component
remains diffuse and extended
because cooling is inefficient (for example the run with adiabatic gas and with added cosmic UV background
presented in Figure 4, see also next paragraph) 
In runs where only radiative cooling is considered,  the longest unstable wavelength is much smaller than the
radius of the residual gas, which implies a laminar viscous timescale close to the Hubble time. This explains
why the gas is not completely stripped in cooling-only runs (see Figure 4).
In conclusion, despite the strongly time-dependent behaviour of all the variables involved in the
problem, both the action and the saturation of stripping seen in the simulations is consistent
with the expected characteristic timescales
and relative importance of the various ram pressure stripping modes.

The next point to discuss in order to understand the different mass loss curves shown in Figure 4 
is the effect of radiation physics/thermodynamics, which we have just hinted 
during the description of the stripping process above.
The temperature evolution determines whether
the gas component stays extended or becomes concentrated in the central,
deeper part of the potential well as a result of tidally induced
bar-driven inflows. Compression from the outer medium heats the gas. 
The gas is rapidly heated
to $10^5$ K, where the cooling function peaks (the initial temperature is
of the gas disk in the dwarf models is in the range $\sim 3000-8000$ K depending
on the initial mass, and is set by hydrostatic equilibrium requirements)
at which point it cools radiatively to $10^4$ K in a fraction of
a dynamical time; such cold gas easily sinks towards the center due
to the torque exerted by the bar and cannot be removed by ram pressure.
Instead, if compressional heating is
not radiated away, as when the gas evolves adiabatically, the increased 
pressure opposes the bar-driven inflow, keeping more gas at larger radii
where it can easily be stripped.  For stripping
to be very effective for dwarfs with initial $V_{peak} \sim 30-60$ km/s it is
sufficient that the  temperature of the gas is kept above $10^4$ K (the
virial temperature of these halos is $3-5 \times 10^4$ K). This
requires some heating source to counteract radiative cooling. 
Mayer et al. (2006) find that for $V_{peak} \sim 30-40$ km/s
such temperature increase can be achieved owing to heating by the (uniform) cosmic UV background at $z > 1.5-2$, 
where the heating rate and ionizing flux rates are
based on standard models for the time evolution of the intensity 
of the cosmic ionizing background (Haardt \& Madau 1996, 2000).

For even higher values of the circular velocity ($V_{peak} \sim 40-60$ km/s)
complete stripping would require an even higher heating rate. This is
likely to be provided by the local ionizing radiation originating from
the primary galaxy (the Milky Way or M31 for the case of the Local Group).
Mashchenko et al. (2004) calculate
that the FUV flux of M31 derived from its $H_{\alpha}$ luminosity 
is higher than the present-day value of the cosmic UV background
out to 10-20 kpc from its center.
Current starbursts have FUV luminosities $10^3-10^4$ higher than the MW and 
M31, in the range $10^{44}-10^{45}$ erg/s (Leitherer et al. 2002).
According to cosmological simulations at $z=2-3$ Milky Way-sized galaxies were
hosting a major central starburst, forming stars at a rate of $\sim 20-30 M_{\odot}$/yr 
(Governato et al. 2007;2009).This is consistent with the star formation rates measured
for $z=2$ galaxies that should evolve into $L_*$ galaxies by the present epoch (Conroy et al. 2007).
Determining the escape fraction of ultraviolet photons that
will contribute to heating and ionizing the gas of the dwarf is a complex task, with the
estimated escape fraction varying between a few percent to more than $10\%$ (e.g. 
Fernandez-Soto et al. (2003), Shapley et al. (2006)).
We choose a simpler approach by assuming that the FUV luminosity of the primary galaxy in the starbursting phase
is comparable to that of a major present-day starburst ($L_{FUV}=10^{45}$
erg/s). At higher redshift dust absorption should be lower since the
metallicity should also be lower, thereby our estimate, while crude, should
be on the conservative side.
Under this hypothesis we obtain that at 30 kpc from the center the
heating rate was 10 times higher than that associated 
with the metagalactic UV background at $z \sim 2$.
We have run new simulations
including such a local, time-dependent UV background 
(the intensity of the flux is modulated by the orbital distance of the
dwarf) and found that ram 
pressure stripping is greatly enhanced, leading to complete
gas removal after a couple of orbits even for $V_{peak} \sim 50$ km/s
(Mayer 2005).
A systematic exploration of the effect of the local ionizing radiation
over a range of mass scales of dwarf galaxies will help assess the overall
relevance of the effect in the context of the evolution of dwarf galaxy satellites.

The implications of these simulations on the enhanced gas stripping resulting when
the ionizing background at high redshift is taken into account are quite
important. Indeed we can argue that if the progenitors of Draco or Ursa Minor fell
into the Milky Way at $z > 1.5-2$ then ram pressure combined with tides was
able to remove their entire gas content in a couple of orbits.
 This translates into a timescale of about 2-3 Gyr after infall assuming a range of orbital times 
consistent with their current distance from the Milky Way. Therefore, in this
scenario the observed truncation of the star formation in Draco and Ursa
Minor more than 10 Gyr ago occurred  as a consequence of the infall of 
these galaxies into the Milky Way halo. 
Draco would have formed most of its stars before infall
since most of the gas becomes ionized while
approaching pericenter for the first time.
This scenario, however, would still be unable to explain the very high
dark matter content of a galaxy like Draco nor why other dSphs, such as Fornax, appear to 
have a similar total mass (baryons plus dark matter) but a much lower dark matter content.
We will see in the next section that these other properties of dSph satellites of the
Milky Way can also be explained once proper assumptions on the structure
of their disky progenitors are made, and the time-dependent nature of the cosmic ionizing background
is taken into account.

\section{A scenario for the origin of classic dwarf spheroidals in a hierarchical Universe;
the timing of infall into the primary}

In Mayer et al. 2007 we built on the results of the previous section and we
proposed a coherent scenario that explains at the same time the origin of the
common properties of dSphs (low gas content, exponential profiles, low luminosity
and surface brightness, low angular momentum content) and their differences (different star formation
histories and mass-to-light ratios). In this model the key parameter is the epoch of accretion
onto the Milky Way and a key assumption is that the progenitors of present-day  classic dSphs were
not simply gas-rich but extremely gas dominated, consistent with what is found in most
present-day dIrrs in the Local Group and elsewhere
in the nearby Universe (Geha et al. 2006). The large gas fractions found in field dwarfs can be
understood in terms of a decreasing star formation efficiency towards decreasing galaxy mass.
Recently, the THINGS survey (Leroy et al. 2008) has confirmed that disky dwarfs have a star
formation efficiency, defined as the fraction of gas (atomic+molecular) that is converted into stars, 
well below that of normal spirals, of order 1\% or less. The low gas surface densities typically
found in dwarfs likely imply a low conversion efficiency between atomic hydrogen and the star forming 
molecular hydrogen phase (Schaye 2004), probably explaining the low star formation efficiency (note that
the conversion conversion between molecular gas and stars does not seem to be lower in
dwarf compared to normal spirals, see Gardan et al. 2007). Such conversion can be even less efficient
in presence of the ionizing ultraviolet flux arising during reionization, which can dissociate
molecular hydrogen (Schaye 2004). Therefore for field dwarfs that were accreted by the Milky Way 
or M31 at $z > 1$ the assumption of mostly gaseous baryonic disk is even more well grounded.

The initial conditions, including the orbits of the satellites, were chosen
based on a hydrodynamical cosmological simulation of the formation of a Milky Way-type galaxy
(Governato et al. 2007).
We found that satellites that were accreted when the cosmic ionizing background was still 
high, roughly before $z=1$, were completely ram pressure stripped of their gas in one to two pericenter
passsages (Figure 6).
As a result their star formation was truncated.
As explained in section 5, the effect of the ionizing radiation is to heat and ionize the gas, making it more
diffuse because of the increased pressure support, and to suppress star formation.
This, in turn, makes it easier to strip even from the
central regions of the dwarf, essentially having the same effect of a reduced binding energy.
Hence gas-rich dwarfs accreted when the UV radiation background was at its peak lost most of their baryonic
content because this was initially in gaseous form, thus naturally ending up with
a very low luminosity. While their baryonic content dropped orders
of magnitude below the cosmic mean as a result of gas stripping, their original central dark matter mass
in the central region around the surviving baryonic core
was largely preserved because dark matter is affected only by tides, not by ram pressure.
This automatically produced very high mass-to-light ratios, of order 100 (see Figure 2 in Mayer
et al. 2007).
We showed that all the final properties of such systems after 10 billion years of evolution, including
the stellar velocity dispersion profiles, resemble those of the classic strongly dark matter dominated
dwarfs such as Draco, Ursa Minor or And IX. Even the brightest among the ultra-faint dwarfs, that have velocity
dispersions $> 5$ km/s (e.g. Ursa Major) may be explained by this model. However, the very low 
characteristic mass scale of most ultra-faint dwarfs suggests that other formation paths might indeed be
more likely (see section 8).
Dwarfs that fell into halos of bright galaxies below $z=1$, when the cosmic
UV radiation dropped by more than an order of magnitude, retained some gas because tides
and ram pressure could not strip it completely, and underwent subsequent episodes of star formation
at pericenter passages due to bar-driven inflows and tidal compression (Mayer et al. 2001a,b).
These ended up in dSphs that are brighter for a given halo mass (or given central
stellar velocity dispersion) compared to the ones that were accreted earlier.
This is should be case with e.g. Fornax, Carina or Leo I. The two regimes of infall epochs 
explain why Fornax and Draco have roughly the same halo peak circular velocities (and thus mass) 
despite having a luminosity and mass-to-light ratio
that differs by about an order of magnitude. Likewise, Carina and Leo I, these being
prototypical cases of dSphs with an extended (episodic) star formation history, have 
a luminosity comparable to Draco but a mass-to-light ratio 5-10 times lower than that of Draco
(Mateo 1998). 

As a final remark, we note that Madau et al. (2008a) argue that a very low efficiency of star formation,
corrresponding to less than $0.1 \%$ of their total mass being converted into stars, would offer 
a solution to the excess in number counts of subhalos with $V_c > 20$ km/s in dark matter-only simulations
(see also Koposov et al. 2009b for a similar interpretation).
Our model for the origin of Local Group dwarf spheroidals provides a clue to
why dSphs were so inefficient at producing a stellar component, thus pointing to
a solution of the substructure problem at the bright end of the mass function of
dwarf satellites of the Milky Way, which essentially contains all the classic dSphs.
Instead of relating the low efficiency of star
formation to photoevaporation and/or suppression of gas accretion by 
the cosmic ultraviolet background,
we argue that it arose naturally from intrinsic low  star formation 
efficiency in the progenitor gas-rich dwarfs (well below 1\%)
combined with copious gas stripping after they were
accreted in the potential of the primary galaxy. Our mechanism is absolutely general in hierarchical
structure formation and should thus apply to dwarf galaxy satellites of any galaxy group.
The combination of an intrinsic low star formation efficiency prior to infall with ram pressure
and tidal stripping can be thought of as an effective feedback mechanism alternative
to reionization and supernovae feedback.
Recent models that relate star formation directly to the formation of molecular hydrogen and
to its dependence on metallicity (Gnedin, Tassis \& Kravtsov 2009; see also Robertson \& Kravtsov
2008) show convincingly that the star formation efficiency was at the low
levels assumed by us, and invoked by Madau et al. (2008a), when the metallicity was below $0.1$
of the solar value. This is because both molecular hydrogen formation and self-shielding from the dissociating 
UV radiation depend on the dust content, which drops with decreasing metallicity.
Interestingly, the {\it current} metallicity  dSphs, both of the ultra-faint ones and of
the classic ones, is comparable or smaller than a tenth of the solar value (Simon \& Geha 2007), 
suggesting that these objects likely
spent a significant fraction of their assembly history at metallicity lower than such
critical threshold. Therefore inefficient star formation played a key role in delivering gas dominated
dwarfs at the time of their infall into the halo of the primary, and at the same time 
is key to understand the missing satellites problem.

\section{The interplay between gas stripping and the tidally-induced transformation
of the stellar component}

The removal of gas due to ram pressure can in principle have effects on
the dynamical evolution of the other components of the dwarf, dark matter
and stars, since it  is initially a non-negligible fraction of the
total mass within the central kiloparsec 
We have performed a number of tests to explore in detail the nature of the
transformation of the baryonic component in the simulations presented
in Mayer et al. (2007), where both stars and gas are present in the initial
disky dwarf.  The main goal was to
determine the relative role of the rapid removal of the gas (Figure 6), which comprises most
of the baryons initially, and that of the repeated tidal shocks in changing
the shape of the stellar component. Although the initial dwarf model is dark
matter dominated by construction, baryons comprise as much as 30\% of the total
mass within the initial exponential disk scale length. Their rapid removal
due to ram pressure stripping might in principle affect the dark matter potential,
possibly lowering its density as found in previous studies where baryonic
stripping occurs as a result of powerful supernovae winds (Navarro, Eke \& Frenk 1996;Gnedin \& Zhao 2002;Read \&
Gilmore 2005).

Figure 7 shows principle axis ratios $s = b/a$ and $q =
c/a$ ($a>b>c$) of stars, calculated from the eigenvalues of a modified inertia tensor (Dubiniski \& Carlberg 1991)
: $I_{ij} = \sum_{\alpha}
x_{i}^{\alpha}x_{j}^{\alpha}/r^{2}_{\alpha}$, where $x_{i}^{\alpha}$
is the $i$ coordinate of the $\alpha$th particle, $r^2_{\alpha} =
(y^{\alpha}_{1})^2 + (y^{\alpha}_{2}/s)^2 + (y^{\alpha}_{3}/q)^2$, and
$y^{\alpha}_{i}$ are coordinates with respect to the principle axes.
We use an iterative algorithm starting with a spherical
configuration ($a=b=c$) and use the results of the previous
iteration to define the principle axes of the next iteration until the
results converge to a fractional difference of $10^{-2}$.
Results are shown for three experiments.
The comparison between the red and the black lines clearly shows that repeated tidal shocks are necessary for
the transformation of the stellar component to occur;
$c/a$ increases by more than a factor of 6 when the dwarf
is continuously tidally shocked relative to a case in which it just loses
most of its baryons owing to ram pressure and then evolves without any
tidal perturbation (in the latter case we removed the primary halo from the
simulation after the gas was stripped).
The green lines in Figure 7 show a test in which about 30\% of the initial gas
content is retained until the end
because the UV background is absent and thus the effect of ram pressure is reduced
(we tested that not including the cosmic UV background at all or including it with the low amplitude expected at
$z < 1$ from the Haardt \& Madau model yields the same result, see Mayer 2005 and Mayer et al. 2006).
The residual gas is funnelled to the center by the bar, where it forms a very dense
knot (Mayer et al. 2006). We notice that despite the fact that this is a small fraction of
the initial
gas content it represents a non-negligible contribution to the central potential because
it all ends up in the inner 0.5 kpc (corresponding to the initial
disk scale length). In this case the dwarf maintains a bar-like, prolate shape
even after 10 Gyr, hence the lower $c/a$.
After 5 Gyr the central density is 8 times higher compared to the simulation
in which complete gas stripping occurs. As a result $t_{shock} > t_{orb}$ within
0.5 kpc, where $t_{shock}$ is $R_{peri}/V_{peri}$, $R_{peri}$ and $V_{peri}$ being, respectively,
the pericentric distance and the velocity at pericenter, and $t_{orb}$ is the orbital time at
0.5 kpc from the center of the dwarf;
the response of the system to the tidal forcing is thus adiabatic instead of
impulsive, hence further morphological changes are inhibited.
This demonstrates that while gas removal does not drive the transformation it represents
a crucial step for the evolution of the
stellar component because the effect of the tidal shocks is considerably weakened when a dense
central gas component is retained.

Figure 7 also shows that $c/a$ does not grow beyond $0.2$ in isolation, even after a few Gyr of evolution.
The average (apparent) ellipticity measured for late-type dIrrs
in the Local Group is larger, $\sim 0.58$, for the 8 dwarfs having a range of rotational velocities comparable with
those of our initial dwarf galaxy model ($V_{peak} \sim 30-40$ km/s) (Mateo 1998). Similar values are found for dwarf irregular
galaxies in clusters (James 1991). However, in comparing with observed values of ellipticity one must take
into account inclination effects, since the $c/a$ discussed for our simulated dwarf is the intrinsic value, i.e.
the value that would be measured by an observer viewing the dwarf exactly edge-on. By viewing the dwarf at
inclinations in the range 30-60 degrees, which statistically are a much more representative situation, the mean apparent
$c/a$ varies in the range $\sim 0.4-0.7$ within 3 disk scale lengths (it is $0.55$ at 45 degrees of inclination), hence
perfectly consistent with the observed values. This indicates that initial conditions used in tidal stirring
experiments are a reasonable match to gas-rich dIrrs.


To summarize, these experiments indicate that the repeated tidal shocks in the impulsive regime
drive the transformation of the shape and mass distribution of the dwarf galaxy,
and that gas removal is a necessary condition for them to be effective. Prolonged
impulsive tidal heating is one of the ways by which tidal stirring can induce the morphological
transformation from disk into spheroid, the other being bar-buckling (see section 3). 
The former is naturally favoured for low mass, low density stellar disks that are 
naturally prone to an impulsive response and also less self-gravitating, while the
latter is favoured at higher stellar densities and thus higher self-gravity since it
is based on the growth of unstable modes (Mayer el. 2001b; Mastropietro et al. 2005).

\section{The ultra-faint dwarfs in the tidal-stirring scenario}

In the last five years deep photometry with the Sloan Digital Sky Survey (SDSS) has unveiled
a new class of ultra-faint dwarfs with structural properties (e.g. low angular momentum, no gas content)
analogous to classic dwarf spheroidals (dSphs) but with masses and luminosities 1-2 orders of magnitude lower (Willman et al. 2005; Belokurov et al. 2007; Simon \& Geha 2007; Koposov et al. 2008)
Indeed with velocity dispersions in the range $2-6$ km/s and luminosities in the range $-2 < M_B < - 7$ these
objects are fainter than most known Globular Clusters, but clearly differ from them because of their
lower stellar surface densities, larger optical radii, and much higher mass-to-light ratios.
The dozen ultra-faint dSphs detected so far all have $M/L \sim 100-1000$, hence they are
are even more dark matter dominated than the darkest among the classic dSphs, such as Draco
and Ursa Minor.  Like most classic dSphs , they are found at distances well within the
virial radius of the Milky Way halo (at galactocentric distances $< 300$) kpc
The brightest of them, such as Ursa Major, may be the product of the same transformation mechanism
elucidated in the previous sections. In this case the initial progenitor managed to form
just a minuscule stellar component before ram pressure aided by the UV background removed most of the
baryons. But the low masses inferred for most of the ultra-faint dwarfs
are below the critical mass threshold for ``surviving'' reionization (Barkana \& Loeb 1999; 
Bullock, Kravtsov \& Weinberg 2001; Somerville 2002;Susa \& Umemura 2004); they would
have been below the threshold even before being accreted by the Milky Way assuming
a factor of 2-3 decrease in their peak velocity dispersion/circular velocity (thus a factor $\sim 8-30$ in mass)  
due to tidal mass loss over billions of years. This implies
that at $z \sim 7-12$,
when reionization began, they could not  hold onto their gas and form stars, rather they
suffered complete photoevaporation. The most plausible explanation for the origin of the ultra-faint
dwarfs is that they formed their stellar components entirely before the onset of reionization (later
gas accretion at $z > 1$ would be difficult with such shallow potential wells). They
are thus the natural candidates for being the "reionization fossils" predicted in the models of
Ricotti \& Gnedin (2005) and Gnedin \& Kravtsov (2006). High resolution cosmological simulations of
galaxy-sized halos indeed find about a thousand subhalos with virial masses $> 10^6 M_{\odot}$ prior to
reionization, hence massive enough for cooling via molecular hydrogen to lead to gas collapse and
star formation before they are accreted onto the primary galaxy (Madau et al. 2008b). 
A low intrinsic efficiency of star formation due to inefficient formation of molecular hydrogen is expected for these low mass, low metallicty systems (see 
section 6). In addition, even modest star formation would strongly affect these systems via supernovae feedback, keeping
the star formation efficiency low (Read et al. 2005).
Inefficient star formation  might thus be the main cause of the extremely low stellar content of the ultra-faint dwarfs.
Moreover, gas photoevaporation owing to the rising ionizing background would remove most of their gas, hence most
of their baryons, leading naturally to very high mass-to-light ratios without
the need of additional environmental processes such as ram pressure and/or tidal stripping.

On the other end, with original masses likely below $10^7 M_{\odot}$, up to two orders of magnitude less than those
of classic dSphs, tidal stirring and tidal mass loss were also very effective in the progenitors of ultra-faint dwarfs.
What we detect today is probably
the surviving inner core of the original dwarf, which responded more adiabatically to the tidal shocks
owing to the high halo concentration/central density expected in very low mass satellites 
(Kazantzidis, Mayer \& Moore 2004; Kazantzidis et al. 2004a).
However, such tiny subhalos were likely never disky in the first place. In fact for standard values of spin parameters 
expected in cosmological halos
their initial angular momentum content was too low to provide more support than thermal pressure 
against the low gravity of their tiny halo potential wells, especially in the presence of the cosmic
ionizing UV background (Kaufmann, Wheeler \& Bullock 2007).
Therefore ultra-faint dwarfs do not owe their current diffuse, spheroidal appearance
and pressure-supported kinematics to the effect of tidal stirring, rather these structural properties
were established before infall.

Finally, an emerging feature of the ultra-faint dwarfs is that some of them appear to be associated with
other classic dSphs such as Sagittarius or Fornax. One example is the very recently discovered 
Segue 2 (Belokurov et al. 2009), where the kinematics support the association with the Sagittarius stream
and with other similarly faint dwarfs, Bootes II and Coma (for Segue 1 recent work does not support the
same association, see Geha et al. (2009)).
It is then natural to postulate that many if not all these ultra-faint dwarfs
are indeed satellites of satellites, the luminous counterparts of subhalos of subhalos seen in the latest generation 
of cosmological simulations resolving scales of tens of parsecs (Madau et al., 2008b).
However, establishing such a link will require further studies  since current cosmological simulations capable of
resolving satellites of satellites do not include the baryons yet. If they were accreted as members
of sub-groups, the ultra-faint dwarfs might
have undergone tidal stirring, and perhaps even ram pressure stripping,  by a larger parent dwarf before 
falling into the Milky Way or M31 halo (Kravtsov et al. 2004; d'Onghia \& Lake 2008).
The first cosmological, hydrodynamical simulations of the formation of low mass galaxies (with masses comparable with 
the likely progenitors of Sagittarius or Fornax) are under way as we write (Governato, Brook, Mayer et al. 2009).
They will be the first step to
study the nature of satellites of dwarf galaxies and assess how plausible is the idea that many of the 
ultra-faint dwarfs are satellites of satellites.

\section{Towards a coherent picture; two populations of dSphs produced by the combined
action of tides, ram pressure and the cosmic ionizing background}

According to the scenario that we presented in this paper a coherent paper is emerging on
the role of environment in shaping the nature of dwarf spheroidal galaxies. In particular,
we posited two main formation paths for dwarf spheroidals, one for the classic dSphs and
one for the ultra-faint dwarfs. Classic dwarf spheroidals live in tidally truncated 
halos with present-day masses of $10^7-10^8 M_{\odot}$, which were typically a factor of 10
more massive, $10^8-10^9 M_{\odot}$, before infall into the halos of the Milky Way
and M31. They were thus born in halo potential
wells too deep for suppression of gas accretion and/or photoevaporation to play a major role
in determining their final properties. Instead, we argue that their current properties,
including their baryonic content and mass-to-light ratio,
were determined mostly by the effect of ram pressure and tides after infall into the
potential of the primary galaxies, with the cosmic UV background still playing some role
in the thermodynamics of the gas and in regulating star formation both before and 
after infall. For classic dSphs two main outcomes are then possible depending on when they
were accreted by the primary halo. At $z > 1$, when the cosmic ionizing UV background 
is more than an order of magnitude higher than today, infalling dwarfs underwent complete
stripping of their gas after one or two orbits and rapid truncation of star formation.
This early infall mode explains an object like Draco, with a star formation history truncated
about 10 billion years ago and a very high mass-to-light ratio.
At $z < 1$, the weaker UV background implies colder and more tightly bound gas in the infalling
dwarfs, stripping is thus less efficient and does not lead to complete gas removal. Instead,
the remaining gas is consumed in bursts of star formation at pericenter passages due to bar-driven
inflows and tidal compression. This late infall mode explains dSphs like Fornax,
with an extended star formation history and moderate mass-to-light ratio.
For both early and late "infallers", we argue that they had 
a disky, rotating stellar component before infall, as in most present-day dIrrs,
that was later turned into a pressure-supported spheroidal
by tidally induced instabilities and tidal heating, a transformation mechanism
that we dubbed tidal stirring.
On the other hand, for ultra-faints dwarfs the formation path is governed by the effect of reionization
, as expected based on their much lower mass scale, which would place them  in halos with 
masses $< 10^7  M_{\odot}$ even prior to 
prolonged tidal mass loss within the primary halo. They were likely born as spheroidal, pressure-supported
objects before reionization, and lost most of their baryons due to photoevaporation thereafter (``reionization
fossils'').
In light of this scenario it is then clear how to interpret the
the anti-correlation between luminosity and $M/L$ (Simon \& Geha 2007), 
which extends all the way from Fornax to Segue 1, one of the faintest newly discovered
dwarfs. This is the result of ram pressure and tidal stripping combined with
the UV background at the bright/high mass end ($\sigma > 6$ km/s, corresponding to an initial $V_{peak} > 20$ km/s before
infall ), and the product of photoevaporation by the UV background 
at the faint/low mass end (($\sigma < 6$ km/s, corresponding to an initial $V_{peak} < 15-20$ km/s before infall).
It remains to demonstrate quantitatively that the efficiency of these various mechanisms for the
removal of the baryons in dwarfs increases with decreasing mass in the way implied by
the observed correlation.

Finally, Kravtsov et al. (2004) have noted how the radial distribution of present-day dSph satellites of the
MW and M31 can be used to constrain different scenarios for their origin. Dwarf spheroidals appear to be more
clustered towards the center of the primary compared to the average subhalos, hence they have a steeper
than average radial distribution profile.
By comparing the distribution of various subsets
of the substructure population of MW-sized halos in cosmological simulations with the one observed for dSphs
they were able to exclude, for example, models in which dSphs are hosted by the most massive substructures
(Stoher et al. 2002),
while they found acceptable models in which dSphs are some of the most massive substructures prior to
infall (this being consistent with the assumption behind all our works, namely that they lost a significant fraction of their mass
since they were accreted).  Strigari et al. (2007) reached analogous conclusions based on the analysis of the mass
function of the most massive substructures.
The question is; would tidal stirring be consistent with the observed radial distribution of dSphs?
This will have to be investigated directly in cosmological simulations with hydrodynamics and star formation
and with enough resolution to study the satellites' population. New simulations obtain
a radial distribution profile of luminous subhalos that is comparable to that of observed dSphs
(Maccio et al. 2009) but they do not have enough resolution to explore the morphologies of
luminous satellites yet, nor they can reliably model hydrodynamical mechanisms such as ram
pressure stripping. However, for the
time being, there is no expected inconsistency. Indeed the assumption that the
dark matter dominated dSphs had to be accreted at $z > 1$ for the morphological transformation to be
effective (Mayer et al. 2007) naturally implies that they were accreted on fairly tight orbits (high orbital
energy, short orbital times); this automatically implies a radial distribution profile steeper than that
of the overall satellites' population.

\section{Caveats, alternatives and future prospects; thick disks, infalling subgroups and dark matter cores}

The tidal stirring simulations carried out so far have always assumed
an initial configuration in which the disky progenitor is constructed
using an equilibrium model with a relatively thin disk of gas and stars.
In reality the stellar disks of dwarfs with $V_c < 40$ km/s are expected to be
rather thick as the gas that formed the disk in
low mass halos reached centrifugal equilibrium with a fairly high
scale height at the mean temperature of $1.5-3 \times 10^4 K$ expected
in the presence of the cosmic ionizing background at $z > 1$ (Kaufmann et
al. 2007). Thermal pressure, as opposed to rotation, would be crucial
to explain the initial equilibrium structure 
of dwarfs embeddeded halos with $V_c < 20$ km/s
that likely hosted the progenitors of ultra-faint dwarfs.
At some level it may also have affected the initial conditions of massive systems that produced the classic
dSphs, and thus should be investigated further with models capable of capturing the
multi-phase structure of the ISM.
A more diffuse, hotter stellar component
argues in favour of a stronger effect of tides, but dynamical instabilities
such as the bar and buckling instabilities invoked in tidal stirring are 
physically associated to an initially cold, rotating stellar configuration. Gas stripping
by ram pressure should instead be enhanced in an initially hotter, more diffuse stellar system
because of the reduced gravitational restoring force of the gas. We are currently investigating the
tidal stirring and ram pressure of a realistic gas-rich dwarf with a turbulent, multi-phase interstellar
medium formed self-consistently in a cosmological simulation (Callegari, Mayer et al., in preparation).

Recently, an alternative model has been proposed according to which some of the classic
dSphs did not fall into the  Milky Way halo as individual satellites
but rather joined the system as part of a subgroup centered around the Large Magellanic Cloud
(D'Onghia \& Lake 2008).
That some satellites are brought inside of the primary halo as part of sub-groups is 
indeed found in cosmological simulations (Kravtsov et al. 2004; Li \& Helmi 2008).
The model is not necessarily in contrast with the general tidal stirring scenario since
the dwarf might have been tidally perturbed significantly by the LMC, which is 1-2 orders
of magnitude more massive than even the brightest among dSphs (Fornax), and then stirred even more
by the Milky Way as the subgroup was broken up by the tides of the latter soon after 
infall. This model would explain the possible alignment of some of the dSphs with the orbits
of the Magellanic Clouds.
It would also imply that a large fraction, if not half of the luminous dSphs, came in as
satellites of the LMC.  This would be naturally explained by how the cooling time
scales with halo mass.  Gas removed by the cosmic ionizing background or by tidal/ram pressure
stripping as the dwarfs were part of the LMC system would have been re-accreted much more easily 
by the subhalos relative to the case in which they would be part of a larger halo such as 
that of the Milky Way. In fact, assuming that such gas would quickly thermalize with the
ambient gas temperature, reaching thus the virial temperature of the primary halo,
the cooling time would be much shorter ($t_{cool} << T_{hubble}$) at the virial 
temperature of the LMC halo, $\sim 10^5$ K.
However, in this model it is still unclear why dSphs have two main modes of star formation history
(truncated or bursty and extended) 
that do not appear to be linked with their being part of the putative LMC sub-group. 
That being said, the potential wells of the dSphs halos might be sufficiently deep to
re-accrete gas while thei are still embedded in groups with the virial masses as that of the LMC halo,
possibly undergoing an extended star formation history.
It will be interesting to assess if
other intermediate mass galaxies in the Local Group, such as M33, massive enough to be a primary
of a subgroup, have  dSphs, surviving or already disrupted into streams, associated to them. Future
surveys of the M31 system will shed light on this possibility.

Previous work has shown that a cuspy dark matter halo fitted by an 
NFW or Moore profile might develop a core as a result of a strong mass outflow
such as that caused by supernovae winds during a burst of star formation (Read et al. 2005).
More recently, the first high resolution cosmological simulations of dwarf galaxy formation have shown that
this mechanism may indeed be efficient at high redshift in the low mass progenitors of present-day gas
rich dwarfs, producing a galaxy with a kiloparsec-scale core at $z=0$ whose slowly rising rotation
curve matches those of typical late-type dwarfs (Governato, Brook, Mayer et al., in prep.).
A clumpy, turbulent interstellar medium may also contribute to core formation via exchange 
of energy and angular momentum with the surrounding dark matter halo (Mashchenko, Wadsley
\& Couchman 2008).
It is thus possible that the progenitors of Local Group dSphs had a cored halo when they accreted onto
the Milky Way and M31. The scenario for baryonic stripping presented in this review is mostly based on
dwarf galaxy models with cuspy (NFW) dark matter halos. Our conclusions regarding the crucial
effect of the environment for dSphs would be even stronger if their progenitors had cores.
Indeed, the various modes of ram pressure stripping discussed earlier are all
enhanced if the gravitational restoring force is lower in the inner region of the dwarf,
as expected in a constant density core. 
Tidal shocks will also be enhanced because the response of the system will be more impulsive with a central core,
rendering the instabilities and the resulting morphological transformation
more efficient. Indeed this was observed in Mayer et al. (2002), where NFW
halos were first introduced in dwarf galaxy models, as opposed to Mayer et al. (2001), where truncated
isothermal halos with a small core of a few hundred parsecs were used.
Other mechanisms that can enhance considerably the intensity of tidal shocks are disk shocking
(see pag. 21-22 and D'Onghia et al. 2009a) and resonances.
D'Onghia et al. (2009b) have recently show that disky satellites that plunge close to the disk of a more
massive companion can undergo a resonant interaction between their rotational motion and their orbital
motion which can heat the stars into a spheroid, along with stripping a large fraction of them, after
only one close pericentric passage. This happens when the spin angular frequency of the disk of the dwarf
and the angular frequency of its orbit are commensurate, and the spin angular momentum and orbital angular 
momentum are nearly aligned (that is, in a nearly prograde encounter).
This mechanism could be an important extension of tidal stirring, perhaps
crucial to explain dSphs orbiting far from the primary, such as Tucana and Cetus in the LG.
Ongoing studies aimed at characterizing in detail the stellar component of these isolated dSphs,
including their star formation history, will soon provide useful constraints on these new
ideas (e.g. Bernard et al. 2009 - see also http://www.iac.es/project/LCID).
Such individual, close encounters may have been more common when the primary halo was still in the
process of assembling from sub-groups and their subhalos were interacting with much larger 
companions, the size of the LMC or so. Some of such encounters may have also resulted into mergers
between dwarfs since the typical velocity dispersions in sub-groups
should have been quite low. Whether this could be another channel of early dSphs formation is
currently under investigation (Klimentowski et al., in preparation). While all these different
tidal mechanisms relying on the idea of pre-processing the morphology of the dwarf satellite
population before they are accreted by the primary
may play a role at some level, they need to be consistent with the morphology-density relation,
a crucial constraint that tidal stirring can naturally satisfy.

As our understanding of dwarf galaxy formation from cosmological initial conditions improves,
interaction simulations will have to be revisited to better calibrate the quantitative effects
of the various mechanisms discussed in this review. While tidal stirring may be only
one part of the overall picture of dSphs formation and evolution, the new studies just recalled reinforce
the idea that tides, in various forms and at different stages of the evolution of satellites, played 
a key role in shaping dSphs as they are today. "Nurture" rather than "nature" is thus the most likely
explanation of the origin of the classic dSphs.

\begin{acknowledgments}
I thank all my main collaborators through the years that made possible to carry out all the work presented in this review; Stelios Kazantzidis, Chiara Mastropietro, Fabio Governato, Ben Moore, Thomas Quinn, James Wadsley, Monica Colpi, Ewa Lokas,
Jaroslaw Klimentowski, Gary Mamon, 
Tobias Kaufmann, Beth Willman and Joachim Stadel, and also PhD students Simone Callegari and Robert Feldmann
for continuing the study of the environmental effects described in this review within the new generation of cosmological
hydrodynamical simulations. I also thank Cristiano Porciani for his help on issues regarding the effects 
of the local ionizing ultraviolet radiation, and George Lake, Justin Read, Nick Gnedin, Oleg Gnedin, Andrey Kravtsov, James Bullock, Piero Madau, Avishai Dekel, Elena D'Onghia, Carme Gallart, Evan Skillman, Pierre-Alain Duc, Julio
Navarro, Arif Babul, Martin Weinberg, Neal Katz, Andrea Ferrara, Wyn Evans,  Gerry Gilmore, Vassily Belokurov , 
Michael Hilker, Pavel Kroupa,
Thorsten Lisker and Jorge Pennarrubia  for interesting and helpful discussions on dwarf galaxies and galaxy interactions over the years.

\end{acknowledgments}

\end{document}